\documentclass[]{JHEP3}
                    
\usepackage{graphicx}
\usepackage{graphics}
\usepackage{epsfig}
\epsfclipon

\usepackage{epsfig,bm}
\usepackage{latexsym}
\usepackage{amssymb,amsmath}

\usepackage{epsfig,bm}
\usepackage{latexsym}
\usepackage{amssymb,amsmath}

\usepackage[small]{caption}

\usepackage{graphicx} 
\usepackage{amsmath}
\usepackage{amsfonts}
\usepackage{amssymb}

\newcommand{\mypicturewidth}{0.8\textwidth}
\newcommand{\halfpicturewidth}{0.4\textwidth}
\newcommand{\thirdpicturewidth}{0.3\textwidth}

\newcommand{\be}{\begin{equation}}
\newcommand{\ee}{\end{equation}}
\newcommand{\bea}{\begin{eqnarray}}
\newcommand{\eea}{\end{eqnarray}}
\newcommand{\mbb}{\mathbb}
\newcommand{\ti}{\times}

\newcommand{\mc}{\mathcal}
\newcommand{\K}{\mc{K}}

\def\lsim{\mathrel{\rlap{\lower4pt\hbox{\hskip1pt$\sim$}}
    \raise1pt\hbox{$<$}}}                
\def\gsim{\mathrel{\rlap{\lower4pt\hbox{\hskip1pt$\sim$}}
    \raise1pt\hbox{$>$}}}                

\title{Preheating After Modular Inflation}

\author{Neil Barnaby$^{1}$, J.~Richard Bond$^{2}$, Zhiqi Huang$^{3}$ and Lev Kofman$^{4}$ \\
Canadian Institute for Theoretical Astrophysics,
University of Toronto, 60 St.\ George St.\, Toronto, Ontario M5S 3H8 Canada \\
$^{1}$barnaby@cita.utoronto.ca, $^{2}$bond@cita.utoronto.ca, $^{3}$zqhuang@astro.utoronto.ca, $^{4}$kofman@cita.utoronto.ca
}

\abstract{ 
We study (p)reheating in modular (closed string) inflationary scenarios, with a special emphasis on K\"ahler moduli/Roulette models.
It is usually assumed that reheating in such models occurs through perturbative decays.  However, we find that there
are very strong non-perturbative preheating decay channels related to the particular shape of the inflaton potential (which is highly
nonlinear and has a very steep minimum).  Preheating after modular inflation, proceeding through a combination of tachyonic instability and 
broad-band parametric  resonance, is perhaps the most violent example of preheating after inflation known in the literature.  
Further, we consider the subsequent transfer of energy to the standard model  sector in scenarios where the standard model particles are confined
to a D7-brane wrapping the inflationary blow-up cycle of the compactification manifold or, more interestingly, a non-inflationary blow-up cycle.
We explicitly identify the decay channels of the inflaton in these two scenarios.  We also consider the case where the inflationary
cycle shrinks to the string scale at the end of inflation; here a field theoretical treatment of reheating is insufficient and one must turn instead
to a stringy description.
We estimate the decay rate of the inflaton and the reheat temperature for various scenarios.
}

\preprint{}


\begin{document}

\section{Introduction} \label{sec:intro}

Reheating at the endpoint of inflation is a crucial requirement for any successful model.
Depending on how inflation ends and how the inflaton interacts, the process of reheating --
the creation of particles from the decaying inflaton and subsequent thermalization -- can proceed differently.
Examples are known of perturbative inflaton reheating \cite{kofman25plus}, non-perturbative preheating \cite{KLS94,KLS97} (leading to excitations of
both bosons and fermions \cite{ferm}), and also reheating via string theory mechanisms such as intermediate
Kaluza-Klein (KK) modes \cite{BBC,KY}, etc.

In the last several years there has been significant  progress in string theory related to the realization of realistic compactifications with stabilized 
moduli \cite{KKLT,BB,BBCQ,CQS}.
This progress has stimulated the development of a new generation of inflationary models based on such stabilized string theory constructions; 
see \cite{MS,Kallosh,Cline} for reviews.
Among the various possible string theory models of inflation, one can distinguish different classes depending on the origin of the inflaton. Modular inflation deals with the inflaton living in the closed string sector.
On the other hand, brane inflation \cite{brane_inflation} deals with that in the open string sector. In the first case it is sufficient to identify one or more 
moduli fields which are already present in the stabilized compactification 
scheme and which are displaced from the minimum of the potential. In the second case, on the top  of the setting required by the
stabilized compactification, the inflaton field must be engineered by
including also probe $D$-branes.  The multitude of possible inflationary scenarios in string theory 
may, at first glance, seem confused and far from unique.  However, this multitude may all in fact be realized if
the paradigm of the string theory landscape is adopted \cite{Susskind}, leading to a picture
where different types of inflation may proceed in different regions of the landscape.
Indeed, even simple field theory models may admit a similar landscape of inflationary possibilities.

String theory inflation models offer a unique opportunity to study (p)reheating in an ultra-violet (UV) complete setting where
it is conceivable to determine all couplings between the inflaton and the standard model (SM) sector from first principles, rather than simply assuming 
some \emph{ad hoc} couplings on a phenomenological basis.  Moreover, because post-inflationary dynamics
are extremely sensitive to the details of these couplings, it follows that any observables generated during preheating 
(such as nongaussianities \cite{ng1,ng2} or gravitational
waves \cite{gw}) have the potential to provide a rare observational window into stringy physics.

Reheating in brane inflation was studied in detail in a number of papers \cite{BBC,KY,creating,sr1,sr2,kklmmt_reheat}.
One of the most interesting realizations of brane inflation is the ``warped''  KKLMMT model \cite{KKLMMT},
constructed in the context of the KKLT stabilized vacuum \cite{KKLT}.  The endpoint of inflation in this scenario is 
the annihilation of a brane-antibrane pair.  (The inhomogeneous dynamics of this
annihilation have been discussed in \cite{caustics,strings}.)  The details of reheating in this model are very complicated
and involve several stages of energy cascading, first from the $D-\bar D$ pair annihilation into closed string modes, next the 
closed string loops decay to excitations of Kaluza-Klein (KK) modes which finally decay to excitations of open string modes on the standard model (SM) brane(s).
An adequate description of this process requires input from string theory.  


Reheating after closed string  inflation, on the other hand, is usually assumed to occur in the regime where ordinary quantum field theory (QFT) is 
applicable.  In this  paper we  investigate in detail the theory of reheating
after closed string modular inflation \cite{CQ,BKPV} models.  We will focus our attention in particular on the scenario of 
K\"ahler moduli \cite{CQ} or Roulette \cite{BKPV} inflation models based on the Large Volume Compactification scheme of \cite{BB,BBCQ,CQS}.  
In this model the role of the inflaton is played by a K\"ahler modulus, $\tau$, (corresponding to 
the volume of a 4-cycle of the internal Calabi Yau manifold) and also by its axionic partner, $\theta$.
However, some of our results may be applicable also in other modular inflation models, such as 
racetrack inflation \cite{racetrack} based on the KKLT compactification \cite{KKLT}.

We find that reheating after  modular inflation can be quite multifarious and may proceed through a variety of different channels (including both 
perturbative and nonperturbative effects and also intrinsically
stringy physics).  The precise identification of these decay channels may depend on model building details, such as the location of the SM in the Calabi Yau (CY) compactification manifold.
In all cases, however, the initial stages of the decay of the inflaton in modular inflation proceed via very strong nonperturbative preheating decay channels.  
This is due to the specific shape of the effective inflaton potential $V(\tau)$, which is very nonlinear and which has a sharp minimum. 
Preheating proceeds through a combination of both tachyonic (spinodal) instability and broad-band parametric resonance and leads to the 
copious production of $\delta \tau$ inhomogeneities (particles).  Within 2-3 oscillations of the background field, nonlinear backreaction effects become
important and the homogeneous condensate is completely destroyed.  In fact, this is perhaps the most violent known example of preheating after inflation! 

In order to understand the full dynamics of reheating in modular inflation we must also identify the decay channels of the inflaton into the visible SM sector.
In the case where the SM is incorporated on a D7 brane wrapped on the inflationary 4-cycle, the K\"ahler modulus $\tau$  decays via a direct coupling to brane-bound SM gauge bosons. 
The initial stages of this decay are nonperturbative and involve parametric resonance of the gauge fields while the later stages involve perturbative decays.
We also consider interactions between the inflaton $\tau$ and brane-bound MSSM fermions such as the Higgsino and gaugino.

On the other hand, the D7 brane construction described above may result in dangerous $g_s$-corrections to the inflaton potential which violate the 
smallness of the slow roll parameters
and spoil inflation \cite{fibre}.  Therefore, it may be desirable to exclude such a wrapping.  In this case the SM can still be localized
on a D7 wrapping some non-inflationary 4-cycle of the CY compactification.  Such a configuration forbids any direct coupling between the 
inflationary
sector and the SM sector, thus complicating the process of reheating.  In this case the inflaton may still decay to SM states via a nontrivial mixing 
between the inflaton fluctuations and the fluctuations of the moduli associated with the 4-cycle that the SM D7 wraps.  The latter may couple directly to brane-bound SM states.
Since the intermediate stages of the inflaton decay involve bulk states, the mixing proceeds via Planck suppressed operators.

Finally, we have identified another reheating mechanism which involves distinctly stringy physics and which does not require a D7 brane to wrap the inflationary 4-cycle.
An important  model  parameter is the value of the inflationary 4-cycles volume, $\tau$, at the stable minimum of the
effective potential, $\tau_m$. If $\tau_{m} > l_s$ then the supergravity description remains valid during both inflation 
and reheating  \cite{CQ}. In this case, reheating treatment is purely field-theoretical and involves calculating
the couplings between the inflaton and standard model degrees of freedom, as described above. However, as was noted  in \cite{BKPV}, 
although the choice of $\tau_m$ does not alter the field-theoretical treatment of inflation, it \emph{does} crucially impact the dynamics of reheating.\footnote{A simple field
theoretic analogue of this scenario is a toy model where the inflaton potential is extremely flat (to provide sufficient inflation) with an extremely steep minimum.  For this toy example
ordinary QFT is valid during the inflationary stage.  However, in the limit that the mass at the minimum approaches $m_s$ the field theoretical treatment of reheating breaks down
and one must instead turn to a stringy description.}  
If $\tau_{min} \le l_s$ then the supergravity approximation is valid \emph{only} during the slow roll (large $\tau$) regime. At the small values of $\tau$
relevant for (p)reheating stringy degrees of freedom, in addition to the supergravity ones,  will be excited.  In this case the physics of reheating 
will change drastically from the naive picture.  One expects, along the lines of \cite{BBC,KY}, that when $\tau$ becomes of order the string scale, light winding modes are created, 
and these subsequently decay into free closed strings which cascade into KK excitations.  These intermediate KK modes can, finally, decay into SM
states on the brane as in \cite{BBC,KY}.  One can also think about the shrinking 4-cycle as an enhanced symmetry point, associated with the quantum production of light
degrees of freedom \cite{beauty}.

This paper is organized as follows.  In section \ref{sec:inflation} we review the large volume compactification of type IIB string theory
and also discuss K\"ahler Moduli/Roulette inflation models embedded within that setting.  In section \ref{sec:oscill} we discuss the decay of
the inflaton in modular inflation via self interactions, showing that fluctuations of the K\"ahler modulus $\tau$ are produced copiously at
the end of inflation via strong preheating effects.  We study this explosive particle production first by solving the linearized equations of motion for the
$\tau$, $\theta$ fluctuations and next by performing fully nonlinear lattice field theory simulations.  In section \ref{sec:SM1} we discuss the decay of the inflaton 
into SM particles (both perturbative and nonperturbative) for the scenario where the SM lives on a D7 wrapping the inflationary 4-cycle.  In section \ref{sec:SM2}
we discuss the decay of the inflaton in SM particles for the case where the SM D7 instead wraps some non-inflationary 4-cycle of the CY.
In section \ref{sec:shrink} we propose another mechanism for reheating in modular inflation which involves distinctly stringy excitations.  Finally, in 
section \ref{sec:conc}, we summarize our results and conclude.  

\section{K\"ahler Moduli/Roulette Inflation in the Large Volume Compactification}\label{sec:inflation}

In this section we briefly review recent progress in constructing stabilized ``large volume'' compactifications in type IIB string theory \cite{BB,BBCQ,CQS} and also describe recent
efforts to embed inflation into such constructions \cite{CQ,BKPV} with the role of the inflaton played by the K\"ahler modulus of the internal CY manifold and its axionic partner.

\subsection{Large Volume Compactification}

Let us first discuss the ``large-volume" moduli stabilization mechanism of \cite{BB,BBCQ,CQS}. 
In this scenario the K\"ahler moduli of the CY manifold are stabilized by both perturbative and non-perturbative effects. As argued in \cite{BB,BBCQ,CQS}, a minimum of
the moduli potential in the corresponding  effective $4d$ theory exists for a large
class of models.\footnote{The only restrictions being the existence of at least one blow-up mode resolving a point-like singularity and also 
the requirement that $h^{1,2}>h^{1,1}>1$, where $h^{1,2}, h^{1,1}$ are the Hodge numbers of the CY.}
An effective $4d$ $\mc{N} = 1$ supergravity is completely specified by
a K\"ahler potential, superpotential and gauge kinetic function.  In
the scalar field sector of the theory the action is 
\be 
\label{full_action}
S_{\mc{N}=1} =
\int d^4 x \sqrt{-g} \left[ \frac{M_P^2}{2} \mc{R} - \K_{,i\bar{j}}
D_\mu \phi^{i} D^\mu \bar{\phi}^j - V(\phi, \bar{\phi})
\right], 
\ee where 
\be
\label{ScalP}
V(\phi, \bar{\phi}) = e^{\K/M_P^2} \left(\K^{i \bar{j}} D_i
\hat{W} D_{\bar{j}} \bar{\hat{W}} - \frac{3}{M_P^2} \hat{W}
\bar{\hat{W}} \right) + \textrm{ D-terms}.  
\ee 
Here $\K$ and $\hat{W}$ are the K\"ahler potential and the
superpotential respectively, $M_P$ is the reduced Planck mass
$M_P = 1/ \sqrt{8 \pi G} = 2.4 \ti 10^{18} \textrm{GeV}$ 
and $\phi^i$ represent all scalar moduli. 

The $\alpha'^3$-corrected K\"ahler potential \cite{BBHL}, after stabilization of the complex structure and dilaton, is
\be
\label{KahlerSimple}
\frac{\mc{K}}{M_P^2} = - 2 \ln \left(\mc{V} + \frac{\xi}{2} \right) + \ln(g_s)
+ \mc{K}_{cs} \,, 
\ee 
Here $\mc{K}_{cs}$ is some constant, $\mc{V}$ is the volume of the CY manifold $M$ in units of
the string length $l_s = 2 \pi \sqrt{\alpha'}$ and $g_s$ is the string coupling. The second term $\xi$  in the logarithm
represents the $\alpha'$-corrections with $\xi = -\frac{\zeta(3)\chi(M)}{2(2\pi)^3}$ proportional to the Euler characteristic $\chi(M)$
of the manifold $M$.  The K\"ahler metric appearing in (\ref{full_action}) was computed explicitly in \cite{BKPV}.

The superpotential depends explicitly upon the K\"ahler moduli $T_i$
when non-perturbative corrections are included 
\be
\label{Super}
\hat{W}   =  \frac{g_s^{\frac{3}{2}} M_P^3}{\sqrt{4 \pi}} 
\left(W_0 + \sum_{i=1}^{h^{1,1}} A_i e^{-a_i T_i} \right) \,.
\ee
Here $W_0$ is the tree level flux-induced superpotential. The
exponential terms $A_i e^{-a_i T_i}$ arise due to non-perturbative
(instanton) effects such as a gaugino condensate on the world-volume 
of a $D7$ brane, or Euclidean $D3$ brane instanton 
(see {\it e.g.} \cite{KKLT, BB,BBCQ,CQS}). 
The K\"ahler moduli are complex, 
\be
\label{T}
T_i = \tau_i + i \theta_i \,, 
\ee 
with $\tau_i$ the volume of the $i$-th 4-cycle and $\theta_i$ its associated axionic partner,
arising from the Ramond-Ramond 4-form $C_4$.  In (\ref{Super}) the constants 
$A_i$, $a_i$ depend upon the specific nature of the dominant non-perturbative
mechanism. For example, $a_i=\frac{2\pi}{g_s}$ for Euclidean D$3$-brane
instantons.

In the simplest cases (such as $\mbb{P}^4_{[1,1,1,6,9]}$)
the volume $\mc{V}$ can be written in terms of the $\tau_i$ as follows: 
\be
\label{vol} 
\mc{V} = \alpha \left(\tau_1^{3/2} - \sum_{i=2}^{n} \lambda_i
\tau_i^{3/2}\right)\,.  
\ee
Here $\alpha$ and $\lambda_i$ are positive constants depending on the
particular model. For example, the two-K\"ahler model with the
orientifold of $\mbb{P}^4_{[1,1,1,6,9]}$ studied
in \cite{BBCQ,CQS,DDF} has $n=2$, $\alpha = 1/9\sqrt{2}$ and $\lambda_2=1$.

The formula (\ref{vol}) suggests a ``Swiss-cheese" picture
of a CY, in which $\tau_1$ describes the 4-cycle of maximal size and
controls the overall size of the CY.  This modulus may be arbitrarily large.
On the other hand, $\tau_2, \ldots, \tau_{n}$, are the blow-up cycles which control the size of
the holes in the CY.   These moduli cannot be larger than the overall size of the compactification
manifold.  This CY manifold is schematically illustrated in Fig.~\ref{fig:CY}.

\EPSFIGURE{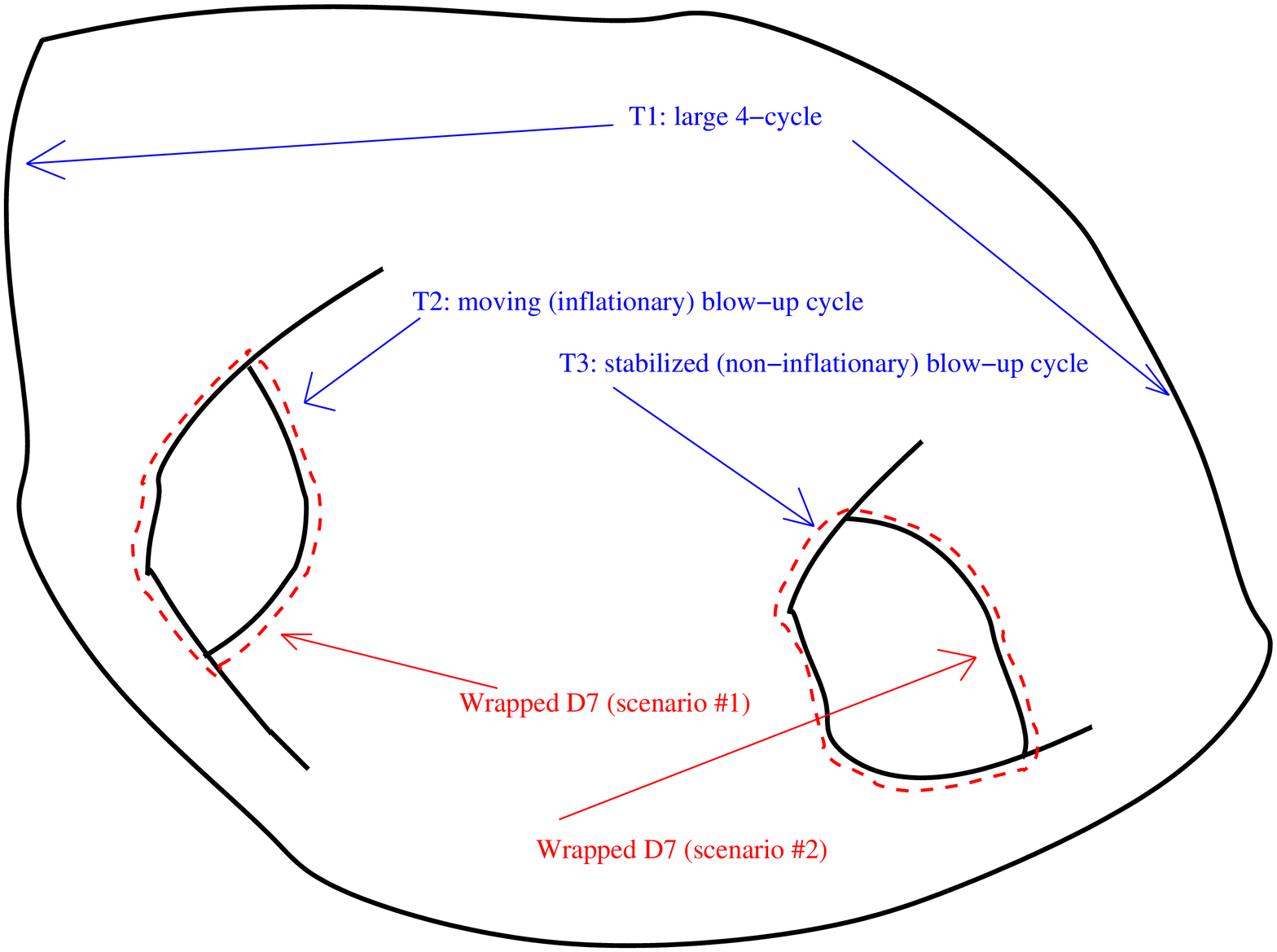,width=5in}{A cartoon of the large volume compactification manifold illustrating the ingredients relevant for K\"ahler 
moduli inflation and (p)reheating.  The modulus $\tau_1 = \mathrm{Re}(T_1)$ 
controls the overall size of the compactification while the moduli $\tau_i=\mathrm{Re}(T_i)$ ($i \geq 2$) control the size of the 
blow-up 4-cycles (hole sizes).  We have labeled
$\tau_2$ as the last 4-cycle to stabilize, and hence this modulus is associated with the final observable phase of inflation.  The total
volume $\tau_1$ and other cycles (for example $\tau_3$) are assumed to already be stabilized.  In the text we consider two possible 
scenarios for the location of the SM: (i) a D7-brane wrapping the inflationary 4-cycle, $\tau_2$,
and, (ii) a D7-brane wrapping the non-inflationary 4-cycle, $\tau_3$.  We have illustrated these wrappings schematically.\label{fig:CY}}

Including both leading perturbative and non-perturbative corrections 
one obtains a potential for the K\"ahler moduli which, in general, has two 
types of minima. The first type is the KKLT minima \cite{KKLT}. 
Relevant for the present study are the ``large-volume" AdS minima studied in \cite{BB,BBCQ,CQS}. These
minima exist in a broad class of models and at arbitrary values of parameters. An important characteristic 
feature of these models is that the stabilized volume of the internal manifold is exponentially large, 
$\mc{V}_{m} \sim \exp{(a\tau_{m} )}$ (here $\mc{V}_m$, $\tau_m$ denote the values
of $\mc{V}$ and $\tau_2$ at the minimum of the potential), and can be
$\mc{O}(10^5-10^{20})$ in string units.  The relation between the
Planck scale and string scale is 
\begin{equation} 
\label{mpl_ms}
  M_P^2 = \frac{4 \pi \mc{V}_{m}}{g_s^2} m_s^2
\end{equation}
Thus, these models can have $m_s$ in the range between
the GUT and TeV scale. 

As argued
in \cite{BBCQ,CQS}, for generic values of the parameters
$W_0$, $A_i$, $a_i$ one finds that $\tau_1 \gg \tau_i$ ($i\geq 2$) and $\mc{V} \gg 1$ 
at the minimum of the effective potential. In other words, the
sizes of the holes are generically much smaller than the overall size
of the CY. 

An important consistency condition for the supergravity approximation to be valid is that
the value of each $\tau_i$ at the minimum of the potential should be larger than
a few.  This criterion ensures that the geometrical sizes of the 4-cycles of the CY are all larger
than the string scale.  One expects any violation of this condition to be associated with the
production of stringy (as opposed to field theoretic) degrees of freedom, such as winding modes.  
More on this later.

\subsection{Roulette Inflation}

Let us now consider inflation in the context of the large volume compactification described above \cite{CQ,BKPV}.  The scenario we have in mind
is the following.  Suppose all the moduli $T_i$ are initially displaced from their minima.  The dynamics will then drive the various fields $T_i$ 
to roll towards their respective minima.  For inflation, we focus on the last modulus $T_2 \equiv \tau + i \theta$ to reach its minimum so that $\tau,\theta$
are still dynamical while all other moduli (in particular the total volume $\mc{V}$) are stabilized.\footnote{The consistency of this approach was discussed in \cite{BKPV},
see \cite{revisited} for more general types of inflationary trajectories.}  See Fig.~\ref{fig:CY} for a cartoon of this scenario.  Note that in the Roulette inflation model \emph{both} 
fields $\tau$ and $\theta$ play an important role in driving inflation.  After having fixed all moduli $T_i$ ($i \not= 2$) we find an effective potential for the inflaton fields $\tau$, $\theta$ of the form   
\be
\label{appr_pot}
V(\tau, \theta)= \frac{g_s^4}{8\pi} \left[ \frac{8 (a_2 A_2)^2 \sqrt{\tau} e^{-2a_2 \tau}}{3\alpha \lambda_2 \mc{V}_m}
+\frac{4 W_0 a_2 A_2 \tau e^{-a_2 \tau} \cos{(a_2 \theta)}}{\mc{V}_m^2}
+\Delta V \right]\ ,
\ee
where we have expanded to order $1/\mc{V}_m^3$ in the (exponentially) small parameter $1/\mc{V}_m$.  In (\ref{appr_pot})
the uplifting $\Delta V$ is some model-dependent constant which should be tuned so that $V=0$ at the minimum.

\EPSFIGURE{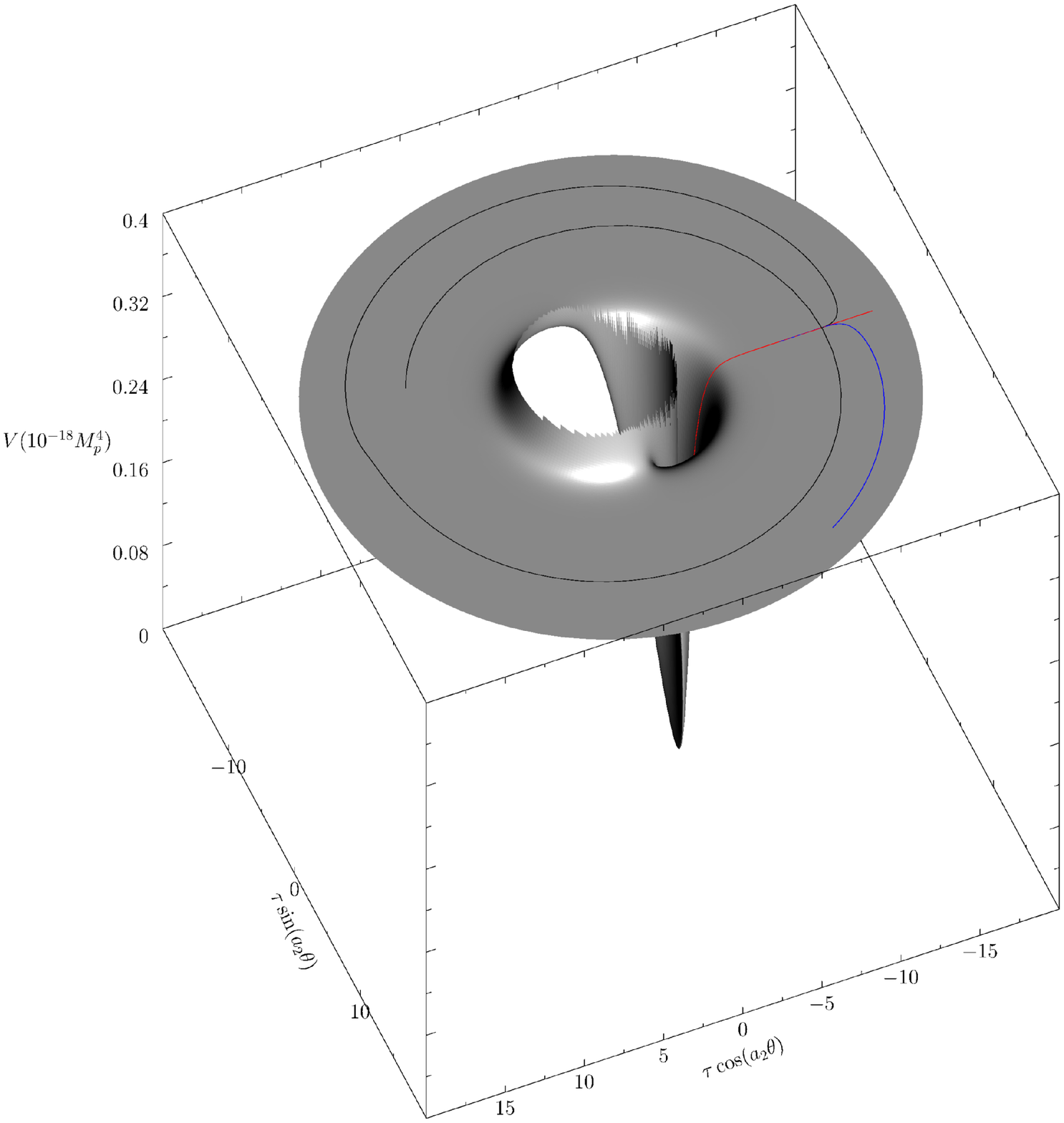,width=5in}{The $T_2$ potential $V(\tau,\theta)$ surface for a representative choice of parameters, using polar coordinates to illustrate the
periodic structure of the potential in the axion direction, $\theta$.  Superimposed on the potential surface are three different inflationary
trajectories, showing the rich set of possibilities in Roulette inflation.  Inflation proceeds in the large $\tau$ region where the potential
is exponentially flat.  On the other hand, preheating after inflation takes place during the phase of oscillations the extremely steep  
minimum near $\tau = \mathcal{O}(1)$ and $\cos(a_2 \theta) = -1$.\label{fig:potential3d_good}}

The potential surface $V(\tau,\theta)$ has a rich structure, illustrated in Fig.~\ref{fig:potential3d_good}. 
The potential is periodic in the axion direction and is exponentially flat (but slightly rippled) at large values 
of the radial coordinate $\tau$ in field space.  The minimum (near $\tau = \mathcal{O}(1)$ and $\cos(a_2 \theta) = -1$)
is extremely steep.  At $\tau \rightarrow 0$ there is a sharp potential barrier, which we have cut-off in Fig.~\ref{fig:potential3d_good}
in order to make the salient features of the potential more transparent. 
The shape of the potential near the edge of the minimum is locally reminiscent of the
 the racetrack inflation potential \cite{racetrack}, as well as the 
``natural inflation'' potential involving a pseudo-Goldstone boson
\cite{natural} (except that in our case both $\theta$ and $\tau$ must be simultaneously
considered).

The potential (\ref{appr_pot}) allows for a rich ensemble of inflationary trajectories in the K\"ahler moduli space \cite{BKPV},
depending on the choice of initial conditions and model parameters.  We have superposed several of these trajectories on 
the potential surface in Fig.~\ref{fig:potential3d_good} for illustration.  Inflationary trajectories may undergo many revolutions
at roughly constant $\tau$, spiraling along the angular direction, $\theta$, like a ball on a roulette wheel.  At some point, the inflaton
eventually gets caught in the trough along $\theta=\frac{\pi (2l+1)}{a_2}$ (where $l$ is an integer) and rolls down to small $\tau$.  Generically the last few e-foldings of inflation
occur along this trough, which is stable against $\delta\theta$ perturbations and along which $V(\tau,\theta)$ takes the form considered in \cite{CQ}.
Inflation ends when the  inflaton rolls into the sharp minimum, like the ball on a roulette wheel falling into the pocket.

Models of this type necessarily introduce a statistical (``gambling'') aspect to the constraints that observational data imposes on inflationary
model building.  In \cite{BKPV} it was advocated to view this theory prior as a probability distribution on an energy landscape with a huge number 
of potential minima.  In \cite{CV} the spectrum and nongaussianity of the curvature fluctuations for the variety of Roulette inflation trajectories was 
discussed.  In \cite{LLPT} cosmological fluctuations were studied more generally in multi-field models with non-standard kinetic terms and the 
Roulette model was considered as a special case.

As discussed previously, consistency of the supergravity approximation 
requires adjusting the parameters so that $\tau_{m} \gsim \mathrm{a}\hspace{1mm}\mathrm{few}$.
We note, however, that even if the SUGRA description in terms of the scalar potential is not valid at
the \emph{minimum}, it still can be valid at large $\tau$, exactly where we wish to realize inflation.  
Hence a small value of $\tau_{m}$ does \emph{not} constrain the  K\"ahler Moduli/Roulette inflation.  Rather, the only consequence of small $\tau_{m}$ 
is that the endpoint of inflation, reheating, would have to be
described by string theory degrees of freedom (rather than simple field theoretic objects). 
In this paper, we will propose one of the possible scenarios of this type.

Finally, we note that there are several types of perturbative corrections that
could modify the classical potential (\ref{appr_pot})
on K\"ahler moduli space: those related to higher string modes, 
or $\alpha'$-corrections (coming
from the higher derivative terms in both bulk and brane
effective actions) and also string loop, or $g_s$-corrections (coming from
closed and open string loop diagrams).  
The most dangerous corrections that could spoil exponential flatness
of the potential at large $\tau$, are the latter category: those coming from open string diagrams. This type of corrections are relevant for the models where the 
SM lives on a D7 brane wrapped on the 4-cycle associated with the inflationary modulus,\footnote{This possibility is illustrated schematically as ``scenario 1'' in Fig.~\ref{fig:CY}.} $\tau_2$, 
and are expected to spoil the flatness of the inflaton potential; see \cite{fibre}
for an estimate of this effect.  One can evade this problem simply by excluding such a wrapping and assuming that the SM lives on a D7 wrapping some \emph{other} (non-inflationary) hole
in the compactification.\footnote{This possibility is illustrated schematically as ``scenario 2'' in Fig.~\ref{fig:CY}.}  This scenario has the disadvantage of complicating
the reheating process, since it forbids a direct coupling between the inflaton and the visible sector.
In this paper, we proceed phenomenologically and consider models both with and without a D7 brane wrapped on the inflationary cycle.  In both cases we identify
the dominant decay channels of the inflaton into SM particles.

It will be handy for the discussion below to have estimates of the masses of the fields
playing a role in inflation and reheating. Recall that from equation (\ref{mpl_ms}) the string scale
and Planck scale are related as $m_s / M_p \sim \mc{V}^{-1/2}$.  The gravitino mass is $m_{3/2} / M_p \sim \mc{V}^{-1}$
and the masses associated with the deformations of the total volume and hole size(s) are, respectively,
given by $m_{\tau_1} / M_p \sim \mc{V}^{-3/2} (\ln \mc{V})^{-1/2}$, $m_{\tau_i} / M_p \sim \ln \mc{V} / \mc{V}$ (for $i \geq 2$).
See \cite{CQastro} for more detailed discussion.

\section{Preheating via Self-interactions in the Inflaton Sector}\label{sec:oscill}

In this section we study the energy transfer from the homogeneous inflaton condensate into fluctuations.  We proceed
by first considering the dynamics of linear perturbations about the homogeneous inflaton background and next by
studying the fully nonlinear dynamics of the system using lattice field theory simulations.  In the next section
we study the subsequent transfer of energy from these fluctuations into excitations of the SM particles.

\subsection{Equations for Linear Fluctuations}

We consider the endpoint of inflation in the Roulette inflation model reviewed in the last section.  To this end we study the action
\begin{equation}
\label{explicit_action}
   S = \int d^4x \sqrt{-g}\left[ \frac{1}{2}\mc{R} - \frac{1}{2}\K_{2\bar{2}} \left( \partial_\mu \tau \partial^\mu \tau + \partial_\mu \theta \partial^\mu\theta   \right) - V(\tau,\theta)   \right]
\end{equation}
where the $(2,\bar{2})$ component of the K\"ahler metric depends only on the 4-cycle volume
\begin{equation}
   \K_{2\bar{2}} = \frac{3\alpha\lambda_2 \left[ 2\mc{V}_m + \xi +6\alpha \lambda_2 \tau^{3/2}   \right]}{2(2\mc{V}_m+\xi)^2\sqrt{\tau}} \cong \frac{3\alpha \lambda_2}{4 \mc{V}_m} \frac{1}{\sqrt{\tau}} + \cdots \label{K22}
\end{equation}
The potential $V(\tau,\theta)$ is given explicitly by (\ref{appr_pot}) and in the second equality of (\ref{K22}) we have expanded to leading order in 
$\mc{V}_m^{-1}$.  Throughout the rest of this paper we use Planck units, $M_P \equiv 1$, although we occasionally write out the factors of $M_p$
explicitly.
For simplicity we restrict ourselves to ``parameter set 1'' (as defined in \cite{BKPV}) corresponding to the choice $W_0=300$, $a_2=2\pi/3$, 
$A_2=0.1$, $\lambda_2=1$, 
$\alpha=1/(9\sqrt{2})$, $\xi=0.5$, $g_s=0.1$ and $\mc{V}=10^6$. We have considered also different parameter choices and found the 
results to be qualitatively similar in all cases.

To understand preheating in the model (\ref{explicit_action}), let us first consider the linearized equations for the fluctuations.  In order to make the 
analysis tractable we introduce the canonical
inflaton field $\phi$ defined by $d\phi = \pm \sqrt{\K_{2\bar{2}}} \, d\tau$ so that 
$\phi \cong \sqrt{\frac{4}{3}\frac{\alpha\lambda_2}{\mc{V}_m}}\, \tau^{3/4}$ in the large volume limit and with some arbitrary convention
for the origin of field space.  Now the action (\ref{explicit_action}) takes the form
\begin{equation}
\label{preheat_action}
  S = -\int d^4x \sqrt{-g} \left[  \frac{1}{2}(\partial\phi)^2 +\frac{1}{2} e^{2b(\phi)}(\partial\theta)^2 + U(\phi,\theta)  \right]
\end{equation}
where
\begin{eqnarray}
  b(\phi) &=& \frac{1}{2}\ln\left[ \K_{2\bar{2}}\left(\tau(\phi)\right)  \right] \\
  U(\phi,\theta) &=& V\left[\tau(\phi),\theta\right]
\end{eqnarray}
Our choice of notation follows \cite{LLPT}.

\begin{figure}
\begin{center}
\includegraphics[width=\halfpicturewidth]{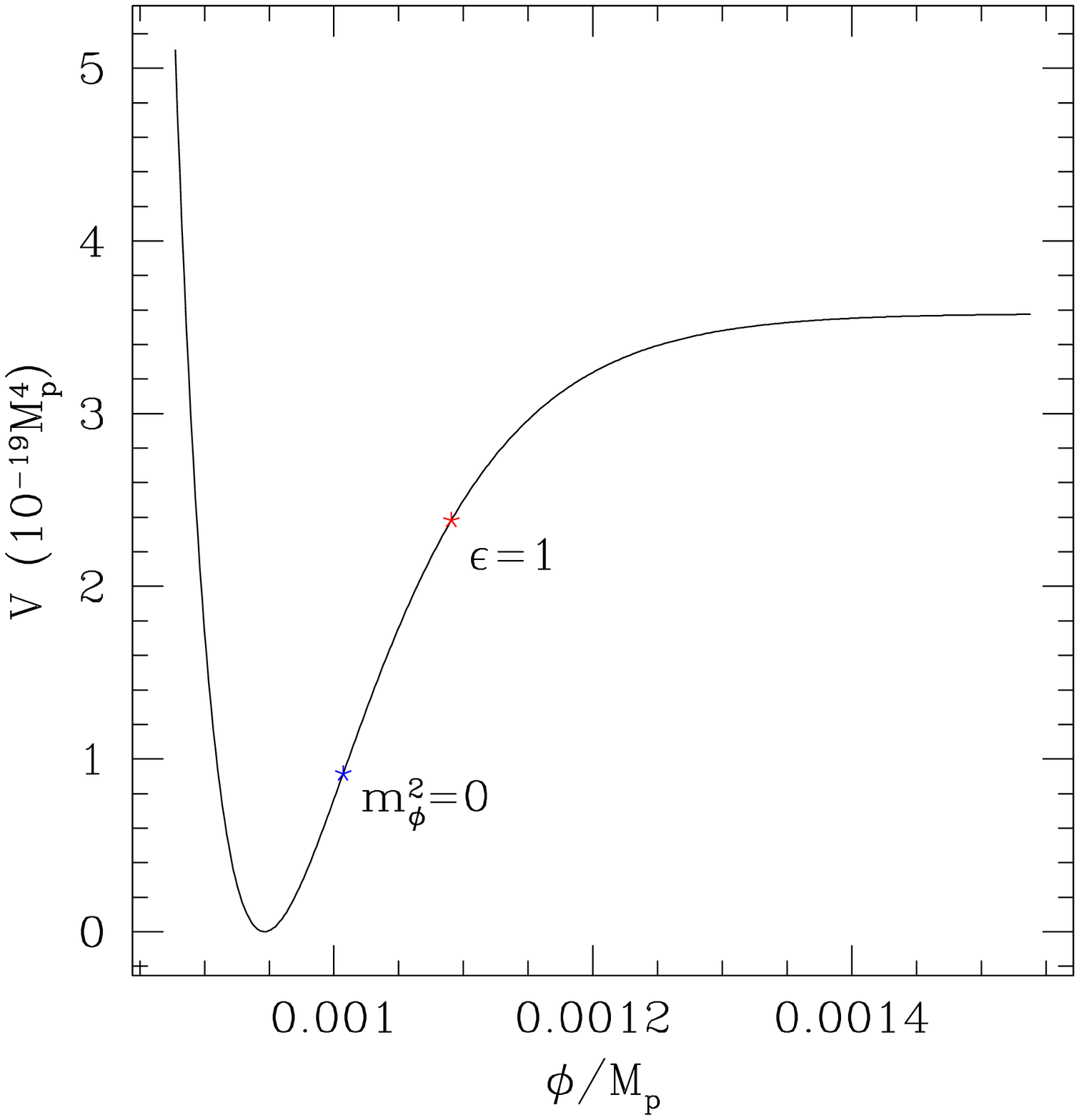}
\includegraphics[width=\halfpicturewidth]{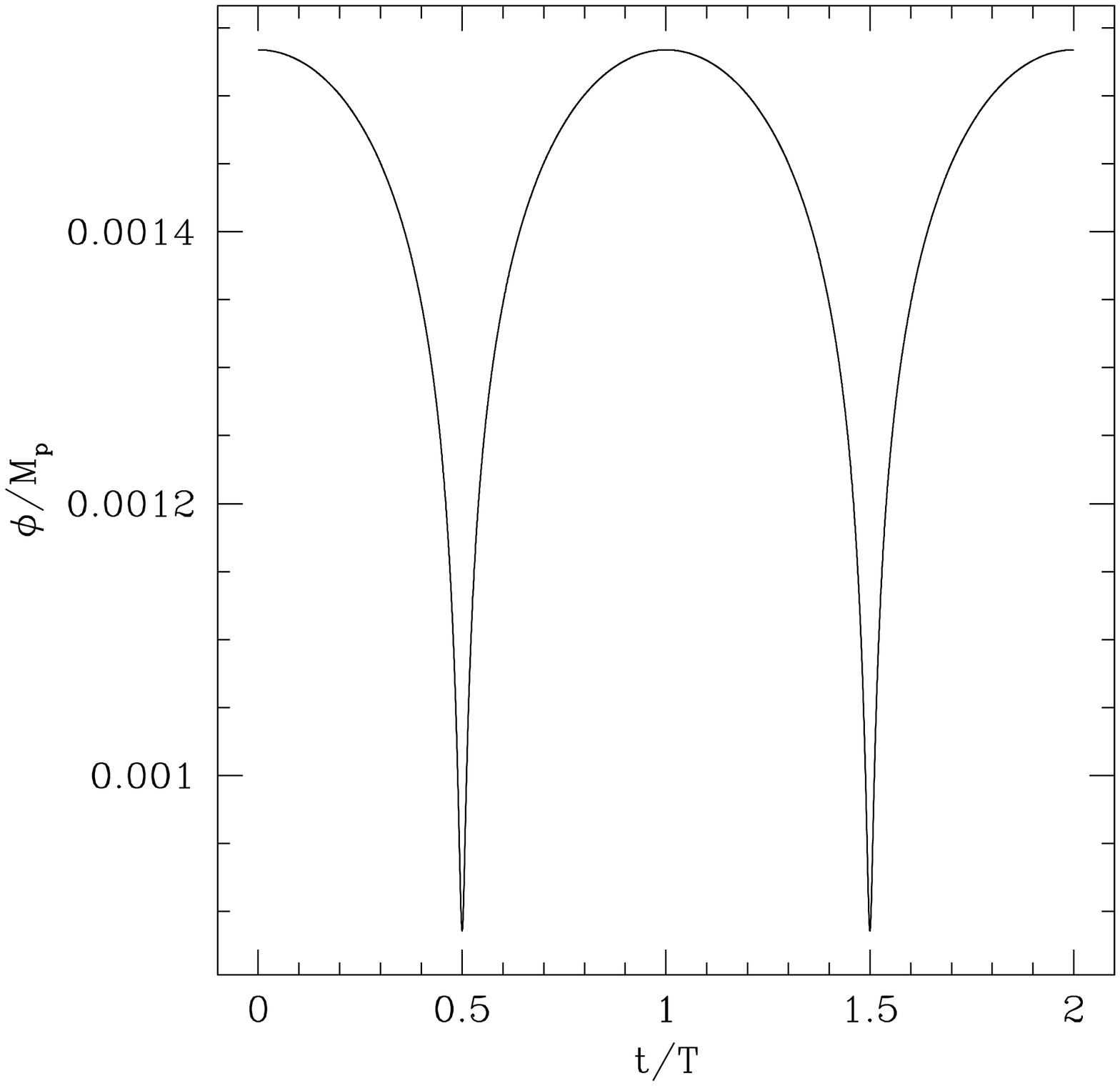}
\caption{The left panel shows the effective potential for the canonical K\"ahler modulus $\phi$ along the axion trough, $U(\phi,(2l+1)\pi / a_2)$ 
   showing the long exponentially flat region relevant for
    inflation and also the steep minimum relevant for the preheating phase of post-inflationary oscillations. 
    We have labeled the point where inflation ends  (where the $\epsilon$ slow roll parameter is unity) and also the point where the effective 
    mass-squared $V''(\phi)$ flips sign, corresponding to the cross-over between the tachyonic and non-tachyonic regions.
    The right panel shows the oscillatory time evolution of the homogeneous inflaton $\phi_0(t)$ at the end of inflation.}
\label{fig:bkg}
\end{center}
\end{figure}

For generic inflationary trajectories, the final few $e$-foldings of inflation take place along the trough $\theta = \pi(2l+1) / a_2$ (with $l$ integer) \cite{BKPV}.  Thus, to study the linear regime of particle production during 
preheating after inflation we expand the fields as
\begin{eqnarray}
  \phi(t,{\bf x}) &=& \phi_0(t) + \delta\phi(t,{\bf x}) \\
  \theta(t,{\bf x}) &=& \frac{(2l+1)\pi}{a_2} + \delta \theta(t,{\bf x})
\end{eqnarray}
In the left panel of Fig.~\ref{fig:bkg} we plot the effective potential for the homogeneous motion of the canonical inflaton $\phi$ along the $\theta = (2l+1)\pi / a_2$ trough, that is $U\left[\phi,(2l+1)\pi / a_2\right]$.  
This potential displays a long flat region at
large $\phi$ which is relevant for inflation and a very steep minimum at $\phi = \phi_m$.  Preheating during inflation takes place during the phase of oscillations about this minimum.  In the right panel of Fig.~\ref{fig:bkg} we plot the time dependence
of the inflaton condensate $\phi_0(t)$ during this oscillatory phase.  Due to the extreme sharpness of the potential minimum the inflaton passes very quickly through the region close to $\phi_m$ while the extreme flatness at larger $\phi$ means that the
inflaton spends a long time near the right-hand side of the valley.  We denote the period of oscillations about the minimum by $T$ and note that for typical parameters $T \sim 10^{-3} H^{-1}$.

Let us now consider the linear fluctuations $\delta\phi$, $\delta\theta$ in the inflaton and axion fields.  The equations of motion for the 
Fourier modes $\delta\phi_k$ and $\delta\theta_k$ take the form
\begin{eqnarray}
  \delta\ddot{\phi}_k + 3H \delta\dot{\phi}_k + \left[\frac{k^2}{a^2} + U_{,\phi\phi} \right]\delta\phi_k &=& 0 \\
  \delta\ddot{\theta}_k + \left[3H + 2b_{,\phi}\right]\delta\dot{\theta}_k + \left[ \frac{k^2}{a^2} + e^{-2b} U_{,\theta\theta} \right]\delta\theta_k &=& 0
\end{eqnarray}
where the quantities $U(\phi,\theta)$ and $b(\phi)$ (and their derivatives) are understood to be evaluated on the unperturbed
values $\phi = \phi_0(t)$ and $\theta=\theta_0(t)\equiv(2l+1)\pi/a_2$.  Neglecting the expansion of the universe\footnote{This neglect is justified since typically the total time for preheating to complete is
of order $10^{-3} H^{-1}$.} and defining the canonical axion fluctuation $\delta\psi$  as
\begin{equation}
  \delta\psi_k \equiv e^{b} \delta\theta_k
\end{equation}
we obtain oscillator-like equations for the mode functions
\begin{eqnarray}
  \delta\ddot{\phi}_k + \omega_{\phi,k}^2 \delta \phi_k &=& 0 \label{phi_mode} \\
  \delta\ddot{\psi}_k + \omega_{\chi,k}^2 \delta \psi_k &=& 0 \label{psi_mode}
\end{eqnarray}
In (\ref{phi_mode},\ref{psi_mode}) the effective frequencies are
\begin{eqnarray}
  \omega_{\phi,k}^2 &=& k^2 + U_{,\phi\phi} \nonumber \\
                    &\equiv& k^2 + M_{\phi,\mathrm{eff}}^2(t) \label{omega_phi} \\
  \omega_{\psi,k}^2 &=& k^2 - b_{,\phi}\ddot{\phi}_0 - (b_{,\phi}^2 + b_{,\phi\phi})\dot{\phi}_0^2 + e^{-2b} U_{,\theta\theta} \nonumber \\
                    &\equiv& k^2 + M_{\psi,\mathrm{eff}}^2(t) \label{omega_psi}
\end{eqnarray}
and we have introduced the notation $M_{\mathrm{eff}}^2(t)$ for the time-dependent effective masses of the fields.
We solve equations (\ref{phi_mode},\ref{psi_mode}) numerically.  As usual \cite{KLS97} we initialize the modes as $\delta\phi_k = 1/ \sqrt{2\omega_{\phi,k}}$, $\delta\dot{\phi}_k = -i \sqrt{\omega_{\phi,k} / 2}$ at $t=0$ (and similarly
for $\delta\psi_k$) which, physically, corresponds to starting with pure quantum vacuum fluctuations with occupation number $n_k =0$.  Violations of the adiabaticity condition $|\dot{\omega}_k| / \omega_k^2 \ll 1$ are associated with 
particle production and lead to the generation of nonzero occupation number, $n_k \not= 0$.  As is typical in the theory of preheating \cite{KLS97} we define the occupation number for the inflaton fluctuations as
\begin{equation}
  n_k^\phi = \frac{\omega_{\phi,k}}{2}\left[ \frac{|\delta\dot{\phi}_k|^2}{\omega_{\phi,k}^2} + |\delta \phi_k|^2  \right] - \frac{1}{2}
\end{equation}
and similarly for the occupation number $n_k^\psi$ associated with the axion.

\subsection{Instability of K\"ahler Modulus Fluctuations}\label{subsec:instability}

Let us first consider the equation (\ref{phi_mode}) for the fluctuations $\delta\phi_k$ of the (canonical) inflaton.  This equation can be viewed as an 
effective Schrodinger-like oscillator equation with ``potential'' determined by the effective
mass term, $M_{\phi,\mathrm{eff}}^2(t)$.  In Fig.~\ref{fig:meff} we plot the behaviour of this effective mass (the dashed blue curve) as a 
function of time.  We see that $M_{\phi,\mathrm{eff}}^2$ has a very particular time dependence: the 
fluctuations $\delta\phi$ are nearly massless during the inflaton oscillations except for periodic ``spikey'' features
uniformly spaced at intervals $\Delta t = T$.  As long as the adiabatic invariant $|\dot{\omega}_k| / \omega_k^2$ remains small, $|\dot{\omega}_k| / \omega_k^2 \ll 1$,
particles are not produced and the occupation number $n_k$ will be close to a constant.
For modes with wave-number $k^2$ sufficiently small, the spikey structure in $M_{\phi,\mathrm{eff}}^2$ leads to extreme 
violations of adiabaticity: $|\dot{\omega}_k| / \omega_k^2 \gg 1$.  In Fig.~\ref{fig:adiabatic} we plot the adiabaticity parameter taking $k=0$ for illustration (the dashed blue curve).  Notice
that each spike in $M^2_{\phi,\mathrm{eff}}$ is also accompanied by a tachyonic phase where $M^2_{\phi,\mathrm{eff}} < 0$.  

\EPSFIGURE{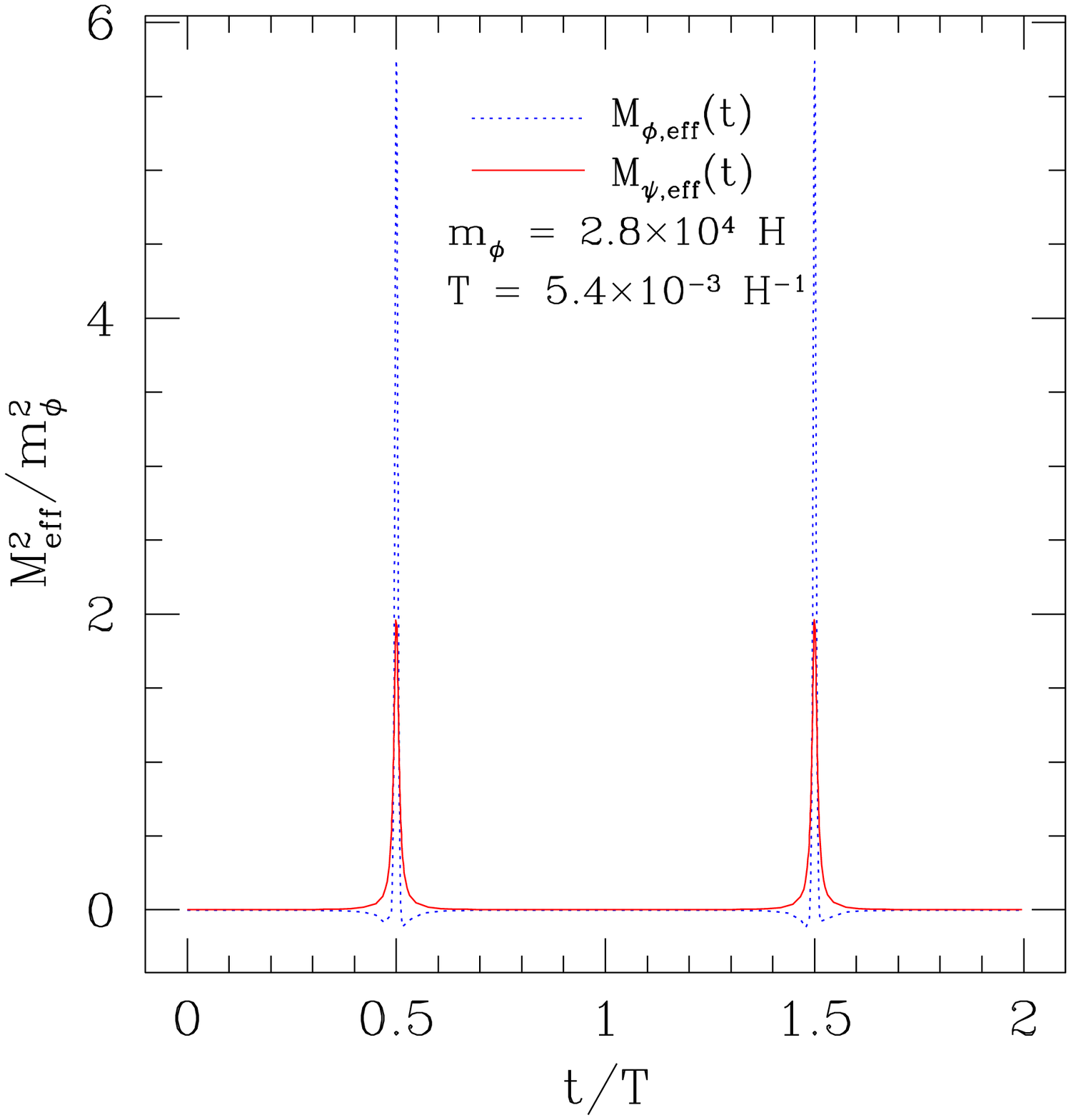,width=\mypicturewidth}{The effective mass-squared of both the K\"ahler modulus $\phi$ (the dashed blue curve) and axion $\psi$ (the solid red curve) as a function of time. 
    We plot the effective mass in units of $m_\phi$, the mass at the minimum of the potential.  Note that 
    $M_{\mathrm{eff}}(t)$
    actually exceeds $m_\phi$ during the inflaton oscillations.  This corresponds to the steep curvature on the left-hand-side of the potential minimum in 
    Fig.~\ref{fig:bkg}.\label{fig:meff}}

\EPSFIGURE{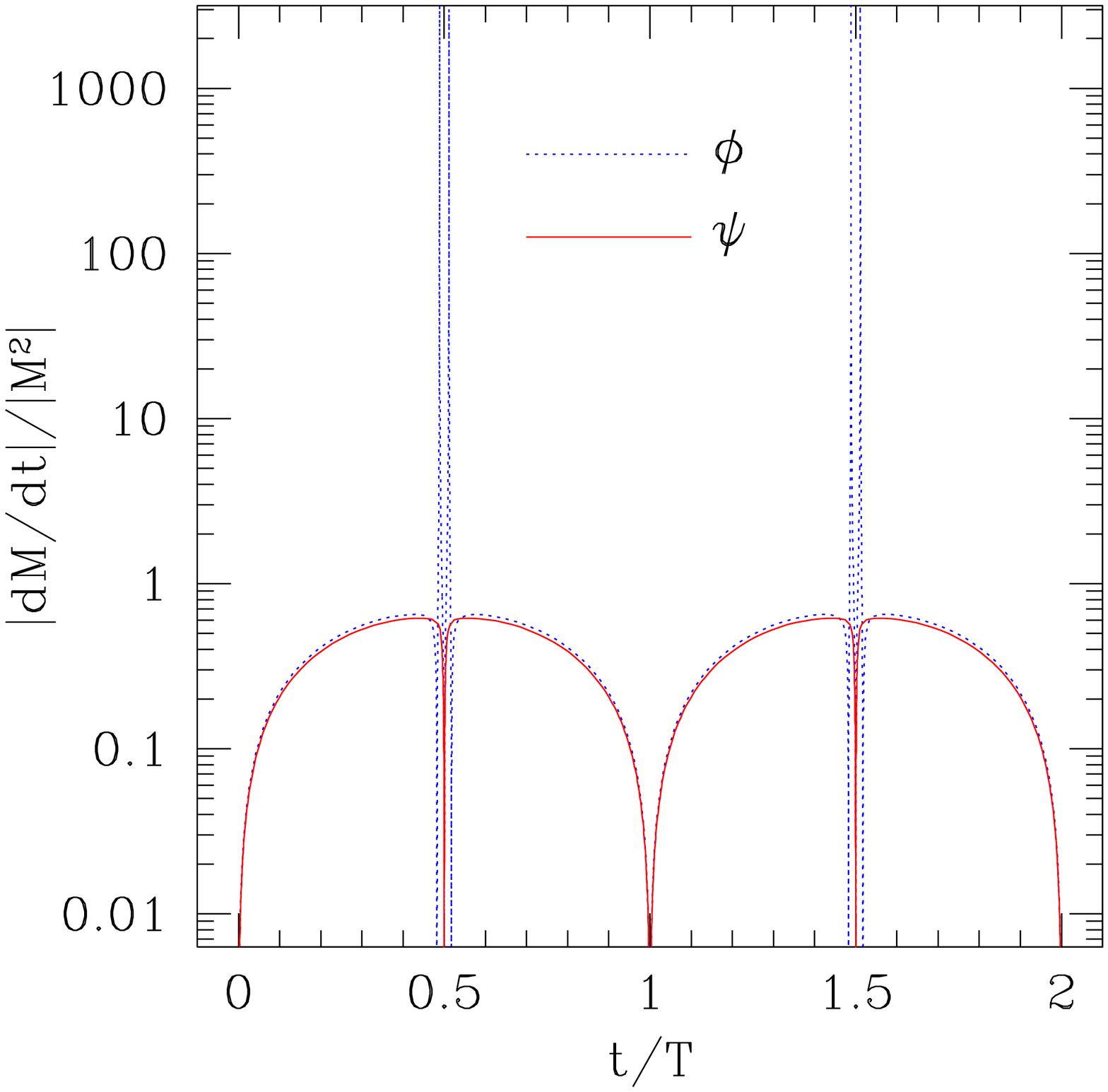,width=\mypicturewidth}{The quantity $|dM_{\mathrm{eff}}/dt|/|M^2_{\mathrm{eff}}|$ 
    for both the K\"ahler modulus $\phi$ (the dashed blue curve) and axion $\psi$ (the solid red curve) as a function of time.  This quantity provides a measure of the violation of adiabaticity and coincides with the adiabatic invariant 
    $\dot{\omega}_k / \omega_k^2$ in the IR.  The spikey structure of the K\"ahler modulus effective mass leads to extremely strong violations of adiabaticity and particle production.  On the other hand, the axion effective frequency varies slowly
    during the inflaton oscillations and axion particles are not produced.\label{fig:adiabatic}}

\begin{figure}
\begin{center}
  \includegraphics[width=\thirdpicturewidth]{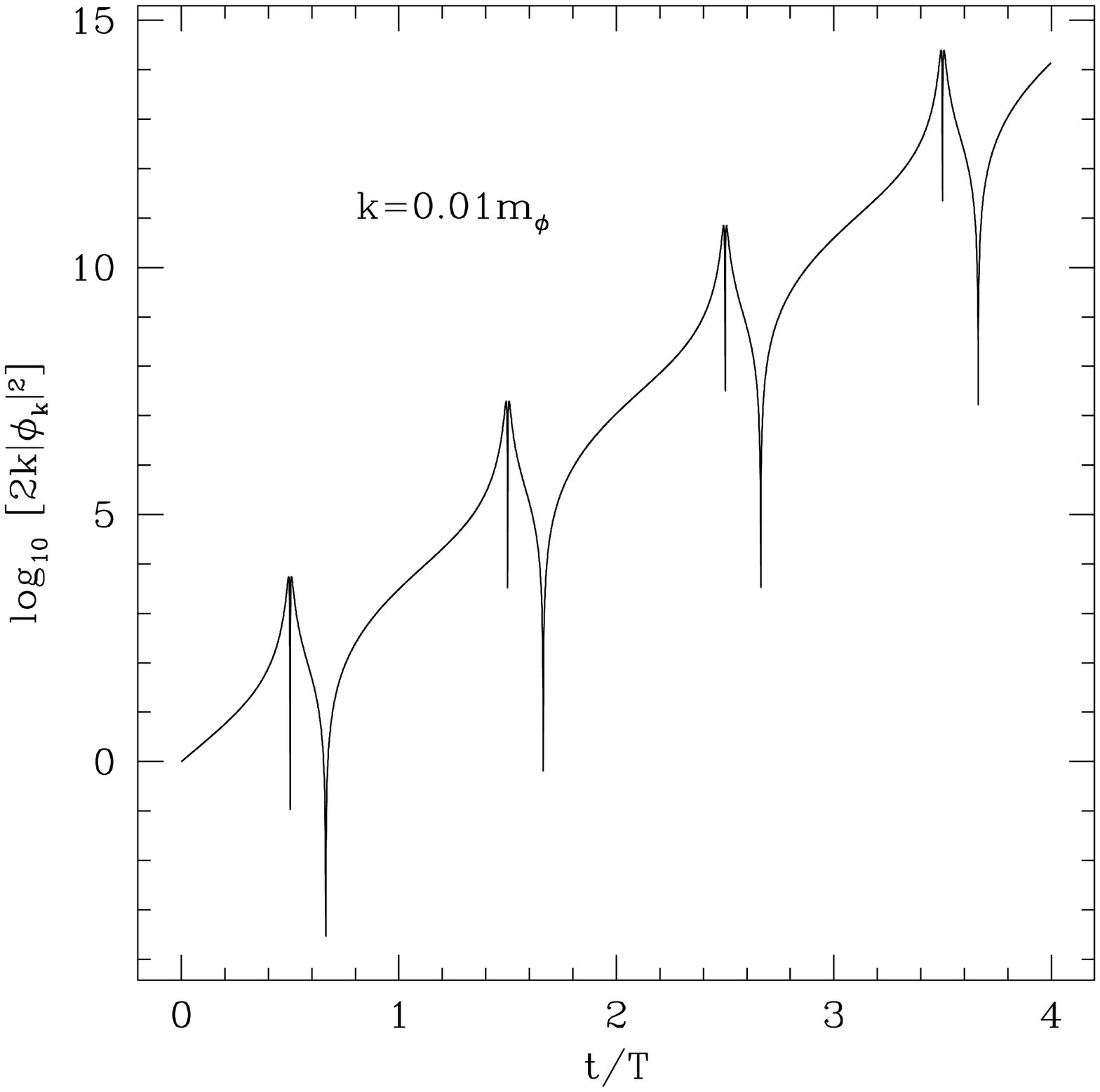}%
  \includegraphics[width=\thirdpicturewidth]{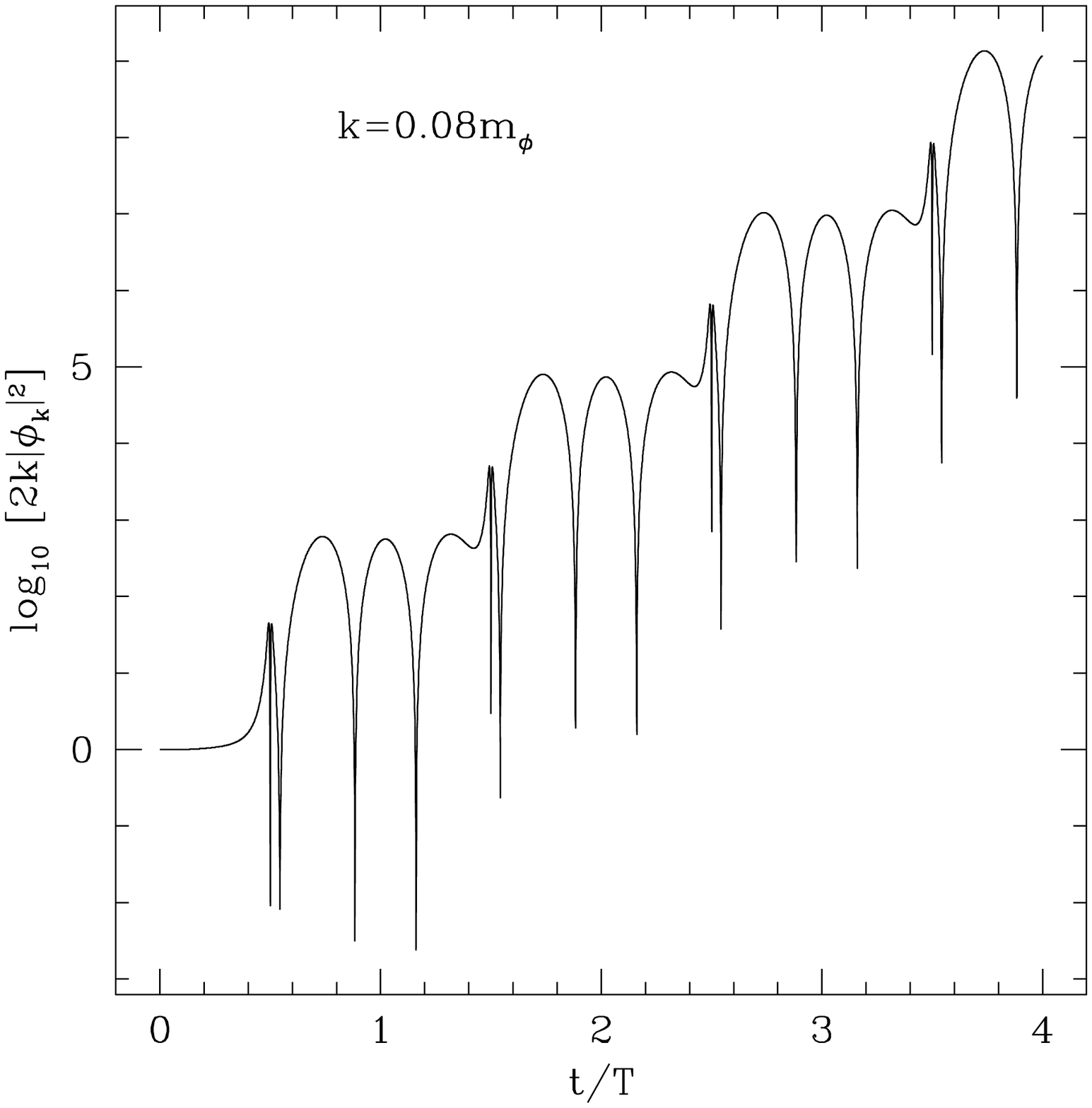}
  \includegraphics[width=\thirdpicturewidth]{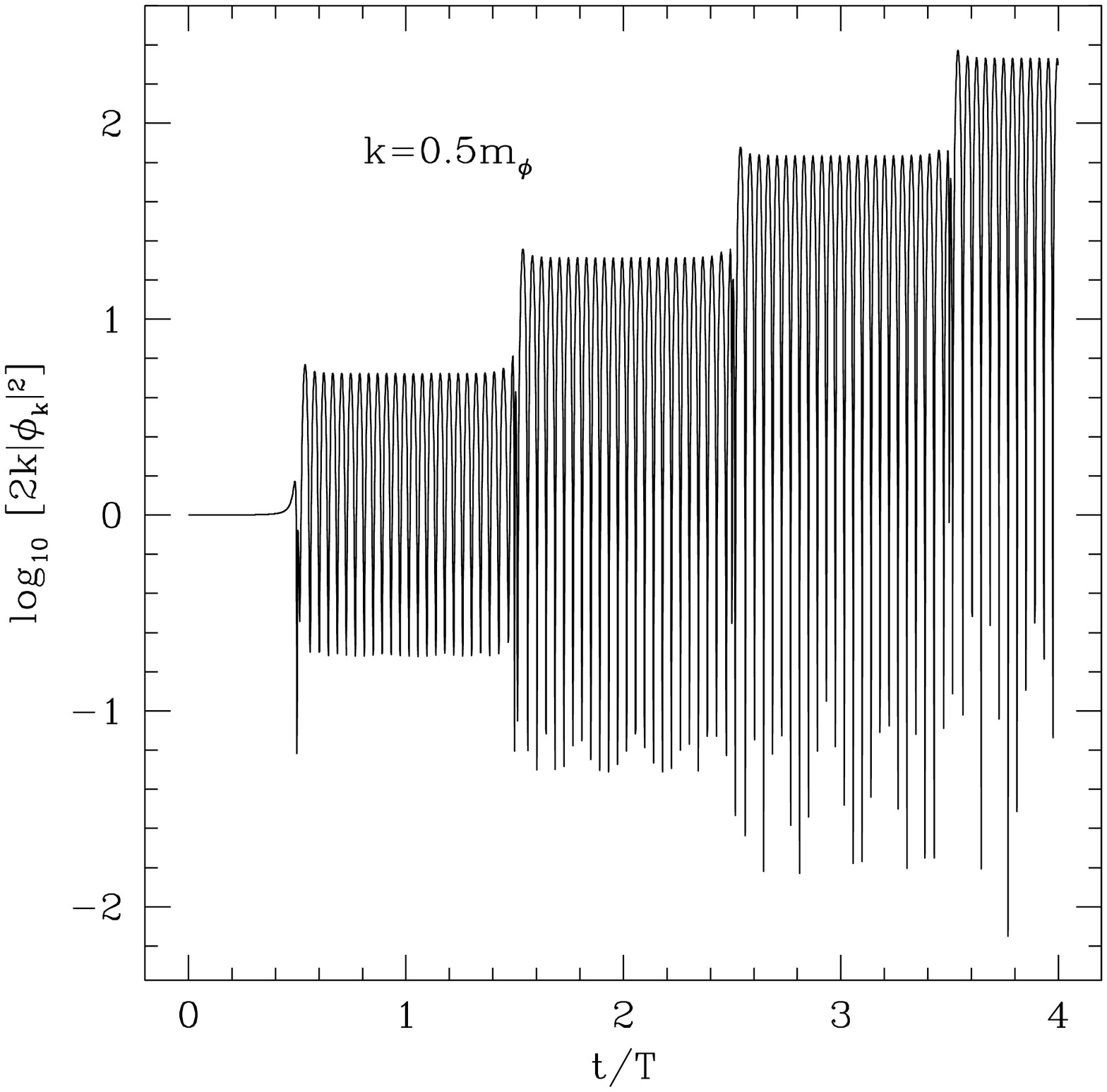}
    \caption{The behaviour of linear K\"ahler fluctuations during preheating after modular inflation illustrating the combination of tachyonic instability and parametric resonance.  The left panel shows the mode behaviour for $k = 0.01 m_\phi$ 
    (corresponding to the IR tachyonic regime) while the right panel shows the mode behaviour for $k = 0.5 m_\phi$ (corresponding to the UV regime of parametric resonance).  The middle panel is $k = 0.08 m_\phi$, corresponding to the intermediate
    regime where both effects are active.}
  \label{fig:modes}
\end{center}
\end{figure}

The periodic time-dependent behaviour of the effective mass $M^2_{\phi,\mathrm{eff}}$ leads to unstable momentum bands where the modes $\delta \phi_k$ grow exponentially as
\begin{equation}
\label{instability}
  \delta \phi_k(t) \sim e^{\mu_k t / T} f_k(t/T)
\end{equation}
with $f_k(t/T)$ some periodic oscillatory function.  We computed the Floquet exponent $\mu_k$ appearing in (\ref{instability})
numerically and the result is displayed in Fig.~\ref{fig:muk_tau}.  The broad unstable region at low $k$ corresponds to tachyonic growth of IR modes coming from the $M^2_{\phi,\mathrm{eff}} < 0$ regions in Fig.~\ref{fig:meff}.
On the other hand, the UV region of Fig.~\ref{fig:muk_tau} featuring a band structure is a result of broad-band parametric resonance.  The modes $\delta\phi_k$ belonging to these two regions of phase space display very different behaviour
and both tachyonic and resonance effects play a crucial role in the dynamics of preheating.  In Fig.~\ref{fig:modes} we plot the time evolution of the K\"ahler modulus mode functions for $k = 0.01\, m_\phi$ (corresponding to the IR tachyonic regime), 
$k = 0.5\, m_\phi$ (corresponding to the UV regime of broad-band parametric resonance) and $k=0.08\,m_\phi$
(corresponding to the intermediate regime where these two effects cannot be disentangled).  Here $m_\phi$ denotes the mass of the canonical inflaton at the 
minimum of the potential.  Note that these different kinds of preheating can \emph{only} be separated in the linear theory; at the nonperturbative
level such a distinction is impossible.

\EPSFIGURE{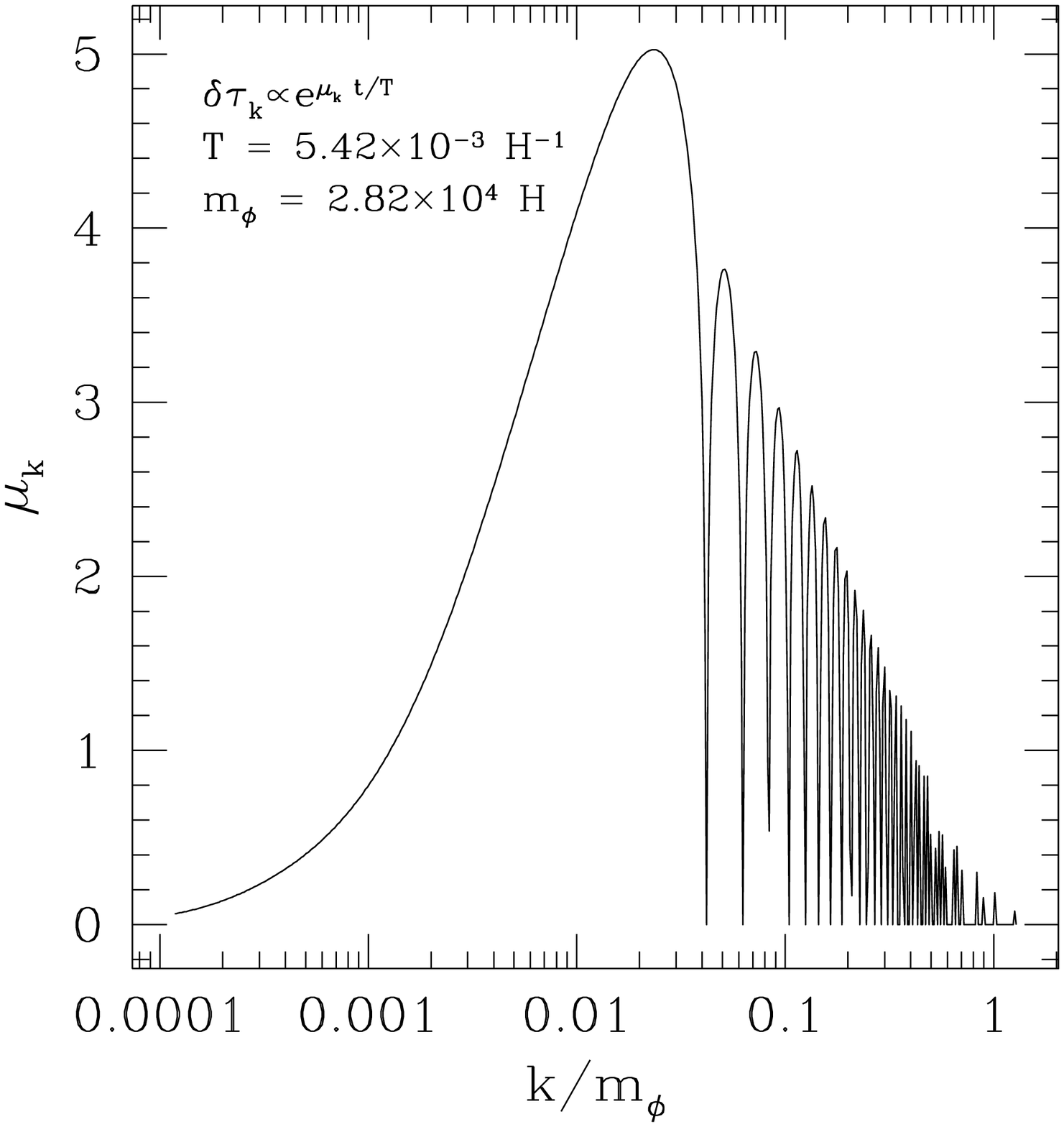,width=\mypicturewidth}{The dimensionless characteristic exponent $\mu_k$ (Floquet exponent) for the exponetial instability of the (canonical) K\"ahler modulus fluctuations $\delta\phi_k$,
defined in (\ref{instability}).  
The broad unstable region in the IR comes from the tachyonic regions 
where $M_{\phi,\mathrm{eff}}^2 < 0$ whereas the UV region displays the band structure that is characteristic of parametric resonance.  The behaviour of 
the modes $\delta\phi_k$ in these two regions is qualitatively different.\label{fig:muk_tau}}

To illustrate how violent the process of preheating in modular inflation is, we consider the energy density $\rho_k$ in a given wave-number $k$,  which 
coincides with $\sim k^4 n_k$ if we evaluate $\rho_k$ at the point in the inflation oscillations where $M_{\phi,\mathrm{eff}} = 0$ (this point is illustrated
on the potential in the left panel of Fig.~\ref{fig:bkg}).  This quantity is plotted as a function of $k$ in Fig.~\ref{fig:k4nk_tau} for three different time
steps in the (linear) evolution.  We see that within only three oscillations of the background field (that is, at $t=3T$) the 
fluctuations contain many orders of magnitude more energy than the condensate.  At this point backreaction becomes critical and the 
linearized treatment breaks down.  Therefore, in order to study the dynamics after the first 2-3 oscillations one must turn to nonlinear lattice simulations, 
which we consider in the next subsection.

\EPSFIGURE{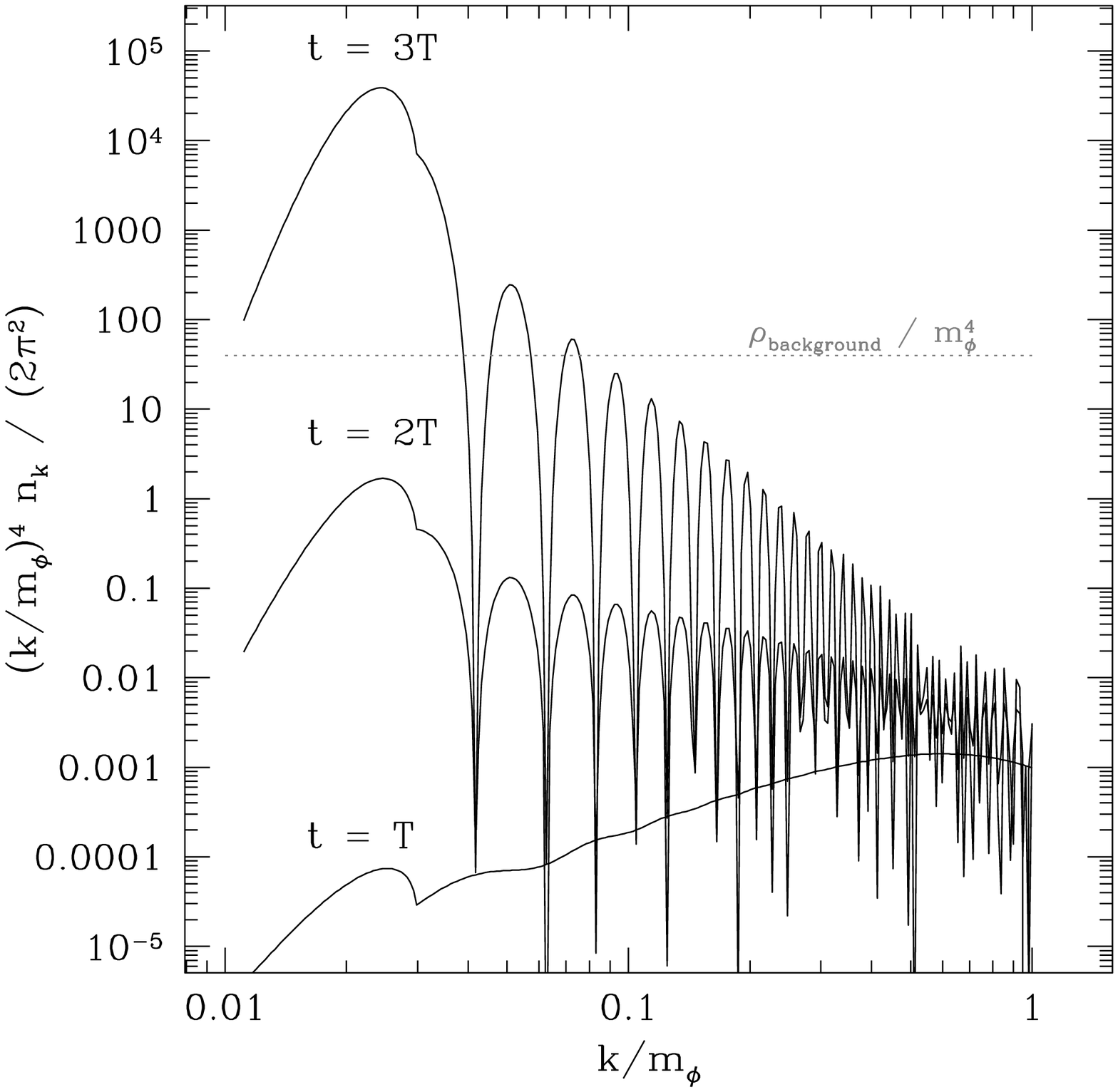,width=\mypicturewidth}{The energy spectrum $\sim k^4n_k$ for $\delta\tau$, calculated with linear theory, showing the rapid and violent production of particles after modular inflation.  Within three oscillations of the background 
field (by $t=3T$) the energy density is significantly larger than the energy in the homogeneous condensate and at this point nonlinear feedback must be taken into account.\label{fig:k4nk_tau}}

We have also studied the dynamics of the linear fluctuations of the axion, equation (\ref{psi_mode}).
The effective mass $M_{\psi,\mathrm{eff}}^2(t)$ for the fluctuations $\delta\psi_k$ of the (canonical) axion.   
is plotted in Fig.~\ref{fig:meff} (the solid red curve).  The behaviour is qualitatively similar to the inflaton effective 
mass, however, the spikes are less sharp and the tachyonic regions are absent.  Thus, we expect the production of axion 
fluctuations to be less efficient.  This intuition is confirmed in Fig.~\ref{fig:adiabatic} where we show that adiabaticity is never 
violated for these modes (see the solid red curve).  Therefore the endpoint of modular inflation is dominated by the extremely 
nonperturbative production of K\"ahler modulus fluctuations, rather than their axionic partner.  During later stages of reheating, 
however, axions may be produced by rescattering effects.

\subsection{Lattice Simulations of Preheating in Roulette Inflation}\label{sec:lattice}

In the last subsection we studied the dynamics of the linear fluctuations $\delta \tau$, $\delta \theta$ at the endpoint of inflation.  
We found that the particular shape of the inflaton potential (which is highly nonlinear and has a very steep minimum) imparts an 
extremely non-adiabatic time dependence to the 4-cycle modulus $\tau$ and leads to explosive production of $\delta \tau$ particles.  
Hence the linearized analysis of this dynamics breaks down very rapidly, within just 2-3 oscillations of the background field, and 
a full treatment requires fully nonlinear lattice field theory simulations.  Hence we now study the evolution of the fields $\tau$, $\theta$ 
in a fully nonlinear way, taking into account also the expansion of the universe self-consistently.

Our numerical simulations are done using a new lattice code which contains (modified) elements of both   
LatticeEasy \cite{latticeasy} and DEFROST \cite{defrost}.  The first version of this code was employed in
the paper \cite{ir} to study infra-red cascading during inflation.  Here we have further generalized the code
to allow for scalar fields with non-canonical kinetic terms.  With the MPI-parallelized code we have used 64
processors on the CITA cluster to evolve the fields $\tau$ and $\theta$ in a cubic box with $512^3$ grids. Vacuum 
mode functions are put in as initial conditions. The very first stage of the evolution, when inhomogeneity
can be treated linearly\footnote{In the case of Roulette inflation this occurs only during inflation and the first oscillation
of preheating.}, is performed in momentum space.  From this initial stage of momentum-space evolution we can determine
which Fourier modes $\tau_k$, $\theta_k$ experience unstable  growth during preheating and ensure that the relevant 
dynamical scales are captured by our simulations (that is, the $k$-modes which dominate the unstable growth are well between
the IR cut-off $k_{IR} = L^{-1}$ defined by the box size $L$ and the UV cut-off $k_{UV} = \Delta x^{-1}$ defined by the lattice spacing 
$\Delta x$).  Once the particle occupation numbers become large we switch over to configuration-space evolution (at the switching
time the gradient energy is roughly $10^{-4}$ as compared to the energy in the homogeneous condensate).  We subsequently
run our configuration-space evolution for a total time $\Delta t = \mathcal{O}(3-4) T$ (where $T$ is the period of oscillations of the
homogeneous background).  This time scale is more than sufficient to see the homogeneous inflaton condensate completely decohered
into inhomogeneous fluctuations.  

To illustrate the rapid and violent development of inhomogeneities during modular preheating we study the energy density
in gradients of the K\"ahler modulus $\tau$, given by
\begin{equation}
  \rho_{\mathrm{grad}} = \K_{,\tau\tau}\frac{1}{a^2}\partial_i \tau \partial^i \tau = \int \frac{d\ln k}{2\pi^2}\frac{k^5}{a^2}\K_{,\tau\tau} |\delta\tau_k|^2
\end{equation}
In Fig.~\ref{fig:k4nk_tau_lattice} we plot the power in gradient energy, given by $k^5 \K_{,\tau\tau} |\delta\tau_k|^2 / (2\pi^2 a^2)$, for three time 
steps in the evolution.  
After the first oscillation, at $t \sim T$, we see the band structure of growing modes in the UV is characteristic of parametric resonance. 
However, by the time $t \sim 2T$ the evolution has become completely nonlinear leading to the destruction of the band structure and the cascading
of power into both the IR and UV.  By the time $t \sim 3 T$ the energy density in gradients is comparable to the homogeneous background energy
density, illustrating the decay of the condensate.  By this time, however, our numerical evolution has become unreliable since the power in 
inhomogeneous fluctuations has cascaded down to the UV cut-off, $k_{\mathrm{UV}} = \Delta x^{-1}$; beyond this point we are bleeding energy into the grid.
In any case, at later times decays into fields which were not included, such as brane-bound gauge fields and fermions, start to
become important.  We consider these decays for a variety of scenarios in the remainder of this paper.

\EPSFIGURE{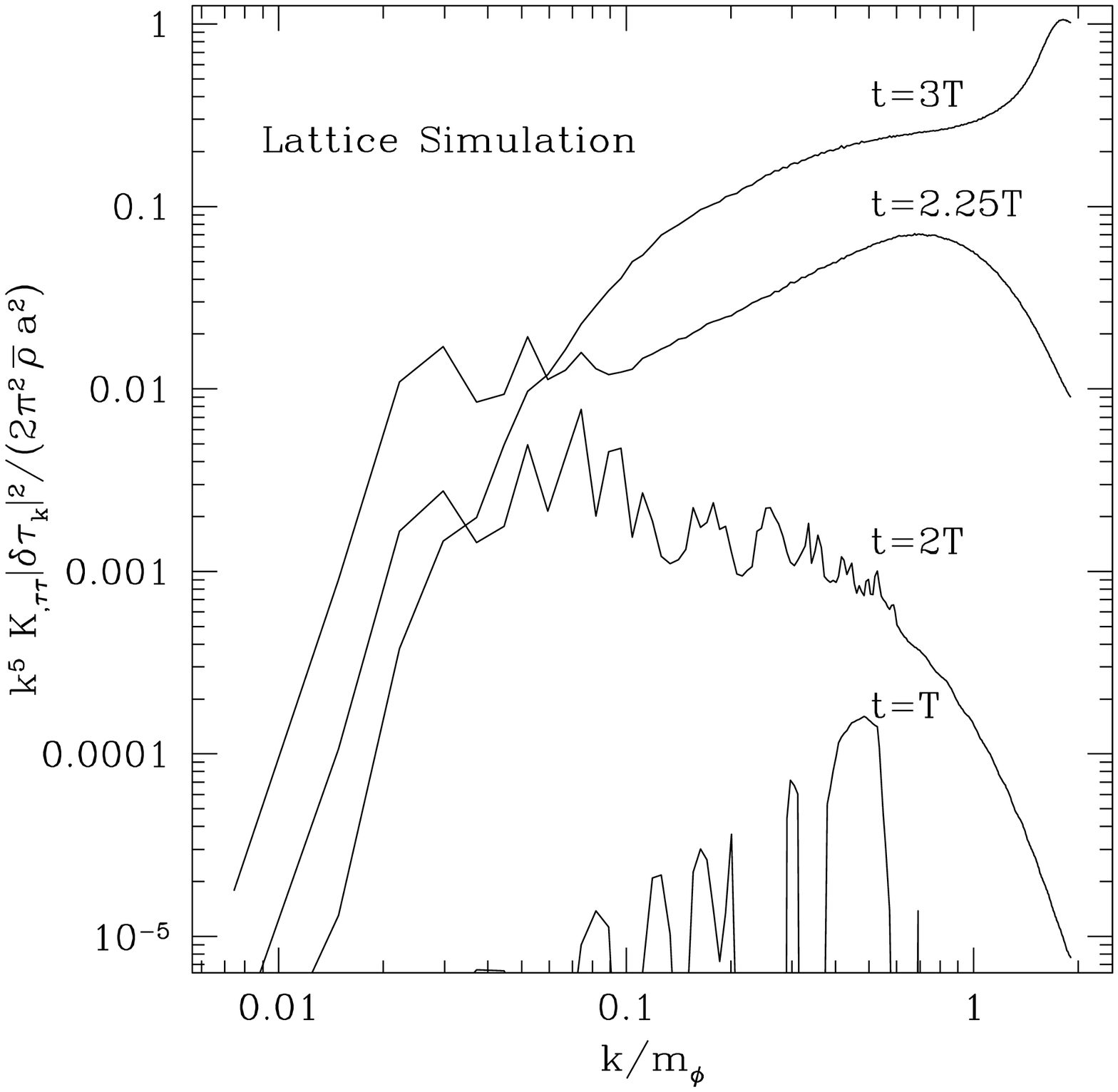,width=\mypicturewidth}{The power in K\"ahler modulus gradient energy, which is given by $k^5 \K_{,\tau\tau} |\delta\tau_k|^2 /(2\pi^2a^2)$, 
    from our lattice simulations, normalized this to the total background energy density, $\bar{\rho}$.  The spectrum is shown for four time steps in the evolution:
    $t=T$, $t=2T$, $t=2.25 T$ and $t=3T$, illustrating the rapid decay of the homogeneous inflaton condensate into inhomogeneous fluctuations.  See the text for further discussion.\label{fig:k4nk_tau_lattice}}

We have also considered the dynamics of the expansion of the universe during preheating.  The effective equation of state during preheating is $\omega = P / \rho \cong 0.1$,
which is very close to a matter dominated expansion.  This is consistent with Fig.~\ref{fig:k4nk_tau} where we see that, in the linear theory,
the $\delta\tau$ particles which dominate the energy density of the universe have momentum in the band $k / m_\phi \sim \mathcal{O}(0.01-0.1)$ and hence we expect these
to be nonrelativistic near the minimum: $\omega_k = \sqrt{k^2 + m_\phi^2} \cong m_\phi + \frac{k^2}{2m_\phi} + \cdots$.  (Note also that the Floquet exponent $\mu_k$ in Fig.~\ref{fig:muk_tau} 
peaks over roughly the same range of scales.)

The violent production of inhomogeneities leads to significant backreaction effects which alter the form of the effective potential, in particular
the location of the minimum and the effective mass at that minimum, $m_\tau$.  The late-time perturbative decays of the produced
inflaton fluctuations $\delta\tau$ into SM fields (which we study in detail in the remainder of this paper) depend sensitively on the value of 
$m_\tau$ at the minimum, hence it is important to determine whether this differs significantly from the value one obtains neglecting backreaction effects.

To illustrate how backreaction effects modify the effective potential, consider a simple scalar field with equation of motion $\Box\phi = V'(\phi)$.
Writing $\phi = \langle \phi \rangle + \delta \phi$ and taking the average of the Klein-Gordon equation we find an equation for the mean field
\begin{eqnarray}
  \Box \langle\phi\rangle &=& \langle V'\left[\langle \phi \rangle + \delta \phi\right] \rangle \nonumber \\
                                        &\equiv& V'_{\mathrm{eff}}\left[\langle \phi\rangle\right]
\end{eqnarray}
In general the effective potential $V_{\mathrm{eff}}$ will differ from $V(\phi)$.  This can be illustrated explicitly by defining the averaging 
procedure as $ \langle V'(\phi) \rangle \equiv  \frac{1}{\sqrt{2\pi}\, \sigma_{\phi}}\int d[\delta\phi] \,  e^{-\frac{\delta\phi^2}{ 2\sigma_{\phi}^2}} V'(\phi)$
with $\sigma_\phi$ the variance of $\phi$.  In general this averaged force differs from $V'(\phi)$.  Note that in our case both inhomogeneous 
fluctuations and nonlinear effects are large, thus a Gaussian distribution for $\delta\tau = \tau - \langle \tau\rangle$ may not be appropriate.  
However, this does not alter the qualitative argument that we are making.

We have estimated the effective mass for the $\tau$ fluctuations at the minimum, $m_\tau$, including backreaction effects, by extracting the rate of oscillation of 
the (spatially) averaged field $\langle \tau \rangle$ from our lattice simulations.  We find that 
\begin{equation}
  m_{\tau}^{\mathrm{new}} = \beta\, m_\tau^{\mathrm{old}}
\end{equation} 
with $\beta \sim 2.5$ for our choice of parameters.  Hence, the true mass including backreaction effects differs from the mass without backreaction by a factor order unity.
However, since the decay of $\delta\tau$ into brane-bound gauge fields depends on the effective mass as $\Gamma \propto m_{\tau}^3$
the value of the factor $\beta$ may be important.

We have also investigated the production of gravitational waves from the rapid development of inhomogeneities associated with the violent
decay of the inflaton after modular inflation.  We have quantified the production of gravitational waves numerically, following the approach
adopted in \cite{gw}.  We did not find any significant production of gravitational waves for the model parameters considered.

\section{Transfer of Energy into the SM Sector: D7 Wrapping the Inflationary 4-Cycle}\label{sec:SM1}

We have seen that the initial stages of preheating in Roulette inflation proceed nonperturbatively by a very strong instability
and lead to explosive production of $\delta \tau$ fluctuations (particles) which destroy the homogeneous inflaton within just a few oscillations.
However, this violent particle production is not sufficient to make contact with the usual hot big bang picture.  We must also ensure that the energy in 
$\delta \tau$
fluctuations can be efficiently transferred into excitations of the visible (SM) sector.  Determining the details of this energy transfer requires a complete 
understanding
of how the inflaton fields $\tau$, $\theta$ couple to SM degrees of freedom and hence this is is necessarily a model dependent issue.  Below we 
will proceed phenomenologically
and consider a variety of possible scenarios, identifying the dominant decay channel of the inflaton in each case.  Throughout this section (and the next) 
we assume that the value of $\tau$ at the minimum, $\tau_{m}$,
is sufficiently large to validate the effective supergravity treatment and hence the details of reheating can be studied using only QFT methods.  In a 
subsequent section we will relax this assumption.

First let us consider the case where the SM lives on a D7 brane wrapping the inflationary 4-cycle, $T_2 = \tau + i\theta$.  This wrapping is illustrated schematically
in Fig.~\ref{fig:CY} as ``scenario 1''.  From the computational point of view this is the simplest imaginable scenario since
it allows for a direct coupling between the inflationary $\tau$ and the fields of the SM.  However, as discussed previously, such couplings lead to 
loop corrections to the inflaton potential 
which are expected to spoil slow roll \cite{fibre}.  Until a robust calculation of such loop effects is available we will proceed phenomenologically and 
suppose that the offending contributions
to the slow roll parameters can be canceled by some fine tuning (or otherwise).

\subsection{Inflaton Coupling to Photons}\label{sec:photon}

Since the inflaton, $\tau$, is the volume of the 4-cycle wrapped by the SM D7 we expect that $\tau$  couples to visible states as an overall prefactor to 
the SM Lagrangian.  Thus, an effective Lagrangian for the inflaton-photon interaction takes the form \cite{CQastro}
\begin{equation}
\label{Lgamma}
  \mathcal{L}_\gamma = -\frac{\lambda}{4}\tau F^{\mu\nu}F_{\mu\nu}
\end{equation}
where $F_{\mu\nu} = \partial_\mu A_\nu -\partial_\nu A_\mu$ is the field strength associated with the gauge field $A_\mu$ and 
$\lambda  = \tau_m^{-1}$ once we have normalized $A_\mu$ so that it has canonical kinetic term when $\tau$ is stabilized at the minimum.

\subsection{Photon Preheating}

\begin{figure}
\begin{center}
  \includegraphics[width=\halfpicturewidth]{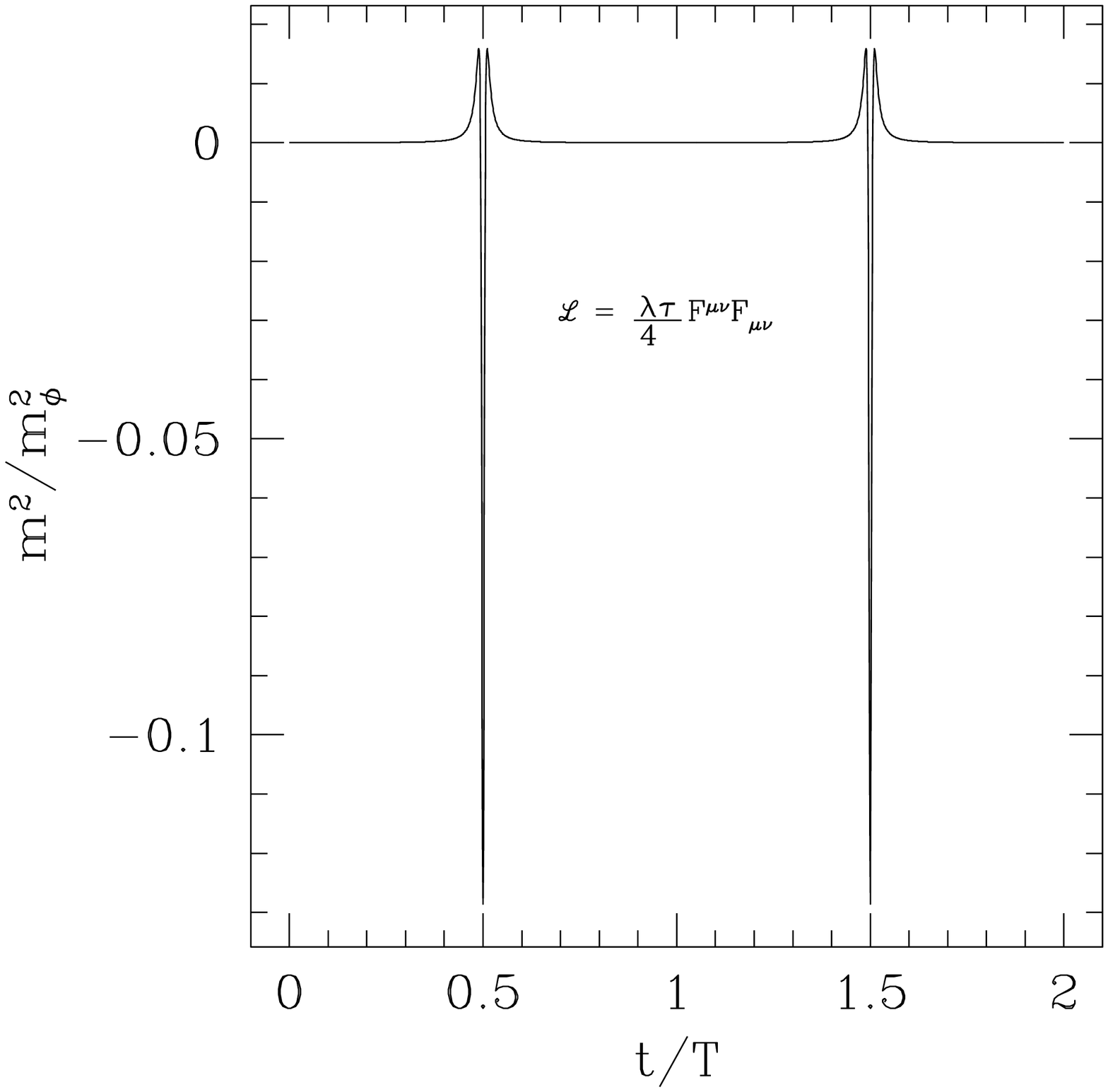}
  \includegraphics[width=\halfpicturewidth]{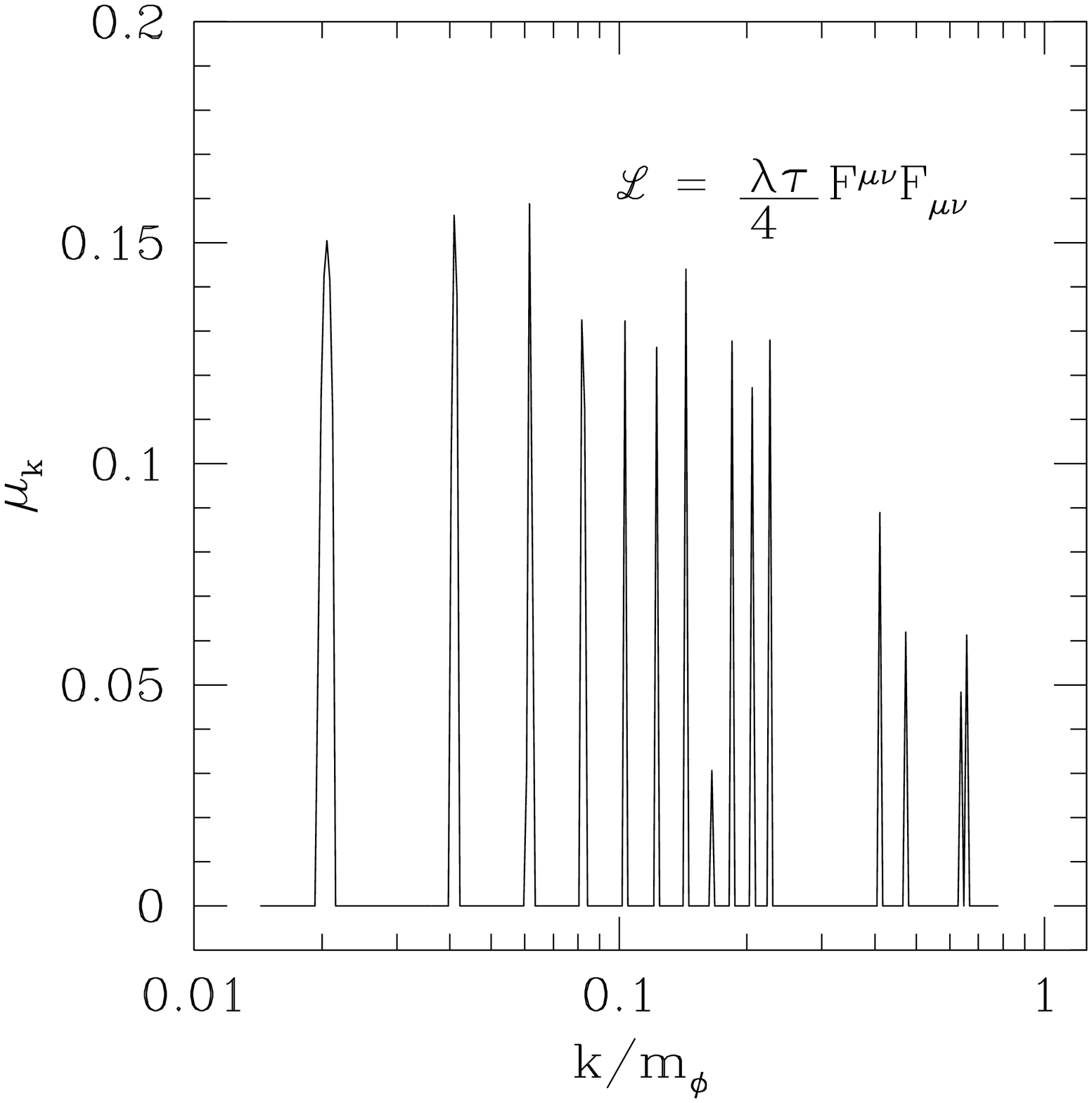}
    \caption{The left panel shows the time dependence of the effective photon mass, $M_{\gamma,\mathrm{eff}}^2(t)$.  
    The right panel shows the Floquet exponent $\mu_k$ for the mode functions $A_k(t) \sim e^{\mu_k t / T}$.}
  \label{fig:gamma}
\end{center}
\end{figure}

The strong nonperturbative production of $\delta\tau$ fluctuations observed in section \ref{sec:oscill} provides a motivation to look for preheating
also in other bosonic fields, such as the photon, $A^\mu$.  Therefore we study the stability of the quantum fluctuations of $A^\mu$ in the background of the
oscillating homogeneous inflaton $\tau_0(t)$.  To this end we choose the transverse gauge: $A^0 = 0$, $\partial_iA^i = 0$.  The equation of motion for the 
Fourier transform of the spatial components of the gauge field $A^i(t,{\bf x})$ in the homogeneous inflaton background
takes the form
\begin{equation}
\label{photon_eom}
  \ddot{A}_k + \left[ H + \frac{\dot{\tau}_0}{\tau_0} \right] \dot{A}_k + \frac{k^2}{a^2} A_k   = 0
\end{equation}
where we suppress the vector index $i$ on the modes $A_k^i(t)$ for ease of presentation.  Equation (\ref{photon_eom}) can be put into oscillator-like form by introducing the field $\tilde{A}_k \equiv (a \tau_0)^{1/2}\, A_k$.  We find
\begin{equation}
\label{photon_oscillator}
  \left[ \frac{d^2}{dt^2} + \omega_{\gamma,k}^2 \right] \tilde{A}_k = 0
\end{equation}
where the time-varying frequency is
\begin{eqnarray}
  \omega_{\gamma,k}^2 &=& \frac{k^2}{a^2} + \frac{1}{4}\left(\frac{\dot{\tau}_0}{\tau_0}\right)^2 - \frac{1}{2}\frac{\ddot{\tau}_0}{\tau_0} - \frac{1}{2} H \frac{\dot{\tau}_0}{\tau_0} - \frac{1}{4}H^2 - \frac{1}{2}\dot{H} \nonumber \\
                      &\equiv& \frac{k^2}{a^2} + M_{\gamma,\mathrm{eff}}^2(t) \label{photon_omega}
\end{eqnarray}
The time dependence of the effective photon mass $M_{\gamma,\mathrm{eff}}^2(t)$ (neglecting the expansion of the universe) is plotted in the left 
panel of Fig.~\ref{fig:gamma}.  The behaviour is qualitatively similar to the effective mass
for the $\delta\tau$ fluctuations (up to an overall sign flip).  The 
oscillatory behaviour of the photon effective mass leads to parametric resonance 
of the photon modes $\tilde{A}_k$ with stability/instability bands in which the photon fluctuations grow exponentially as 
$\tilde{A}_k(t) \sim e^{\mu_k t / T}$.  We have computed the Floquet exponent, $\mu_k$, numerically and this result is plotted in 
the right panel of Fig.~\ref{fig:gamma} where we see the characteristic features of narrow band parametric resonance.

We have seen that very early stages of oscillations of the inflaton condensate $\tau_0(t)$ leads to copious production of photons living on the SM 
D7-brane via parametric resonance preheating.  This phase of violent, nonperturbative photon production
is very short-lived since the production of $\delta\tau$ inhomogeneities completely destroys the condensate within 2-3 oscillations.  The later stages
of the production of SM photons will instead involving perturbative decays $\tau \rightarrow \gamma\gamma$, to which we now turn our attention.

\subsection{Perturbative Decays to Photons}\label{subsec:pertphotons}

After the phase of violent nonperturbative particle production discussed above the homogeneous inflaton $\tau$ will settle down to the minimum
of the potential and reheating will be dominated by the perturbative decays of produced
inflaton fluctuations into SM photons.    To study such processes let us consider the Lagrangian for the total volume modulus $\tau_1\equiv \tau_b$
(where the subscript $b$ stands for ``big'') and the hole-size $\tau_2\equiv\tau\equiv\tau_s$ (where the subscript $s$ stands for ``small'').  Writing
$\tau_i = \langle\tau_i\rangle + \delta \tau_i$ (with $i = b,s$) we have, near the vicinity of the minimum
\begin{equation}
\label{pert_L}
  \mathcal{L} = -\K_{i \bar{j}} \partial_\mu (\delta\tau_i) \partial^\mu (\delta\tau_j) - V_0 - (M^2)_{ij}(\delta\tau_i)(\delta\tau_j) 
     - \frac{\lambda}{4}\,\tau_s\, F^{\mu\nu}F_{\mu\nu} + \cdots
\end{equation}
The mass matrix $(\K^{-1} M^2)_{ij}$ is not diagonal, nor are the fields $\delta\tau_i$ canonically normalized.  Following \cite{CQastro} we
can put this action into a more conventional form by introducing the physical modulus fields $\Phi$, $\chi$ defined by
\begin{equation}
  \left(\begin{array}{c}
    \delta\tau_b  \\
    \delta\tau_s 
  \end{array}\right)
  = \left(\begin{array}{cc}
       R_{b\Phi} & R_{b \chi} \\
       R_{s\Phi} & R_{s \chi}
    \end{array}\right)
    \left(\begin{array}{c}
    \Phi \\
    \chi
  \end{array}\right)
\label{massmatrixmixing}
\end{equation}
The elements of the rotation matrix $R_{ij}$ are computed explicitly in \cite{CQastro}, however, parametrically we have 
$R_{b\Phi} \sim \mc{V}^{1/6}$, $R_{b\chi} \sim \mc{V}^{2/3}$, $R_{s\Phi} \sim \mc{V}^{1/2}$ and $R_{s\chi} \sim \mc{V}^0$.  
Thus the canonical field $\Phi$ is mostly the small 4-cycle $\delta \tau_s$, while the field $\chi$ is mostly 
the big 4-cycle $\delta \tau_b$.  In terms of the canonical basis $\chi,\Phi$ the Lagrangian (\ref{pert_L}) takes the form
\begin{eqnarray}
\mathcal{L} &=& -\frac{1}{2}(\partial\Phi)^2 - \frac{1}{2}(\partial\chi)^2 - V_0 - \frac{m_\Phi^2}{2}\Phi^2 - \frac{m_\chi^2}{2}\chi^2 
                     - \frac{1}{4}F^{\mu\nu}F_{\mu\nu} \nonumber \\
            &+& \frac{\lambda_{\chi\gamma\gamma}}{4}\, \chi\, F^{\mu\nu}F_{\mu\nu} +
                     \frac{\lambda_{\Phi\gamma\gamma}}{4} \,\Phi \,F^{\mu\nu}F_{\mu\nu} \nonumber \\
            &+& \lambda_{\chi\Phi\Phi} \chi \Phi^2 + \lambda_{\Phi\chi\chi}\Phi \chi^2 + \lambda_{\chi\chi\chi}\chi^3 + \lambda_{\Phi\Phi\Phi}\Phi^3 +   \cdots \label{canbasis}
\end{eqnarray}
The masses of the canonical moduli are  
\begin{eqnarray}
  m_\chi &\sim& \frac{1}{\mc{V}^{3/2} (\ln \mc{V})^{1/2}} M_p \label{mchi}\\
  m_\Phi &\sim&  \frac{\ln \mc{V}}{\mc{V}} M_p \label{mPhi}
\end{eqnarray}
so that $\Phi$ is significantly more massive than $\chi$, consistent with the intuition that since $\tau_b \gg \tau_s$ the volume 
modulus should be lighter than the modulus associated with the hole size (at the minimum).  The light $\chi$ particles have Planck suppressed
couplings to brane-bound gauge fields  
\begin{equation}
\label{lambdachi}
  \lambda_{\chi\gamma\gamma} \sim \frac{1}{M_p\, \ln \mc{V}}
\end{equation}
since the fluctuations of the
total volume are bulk modes.  On the other hand, the coupling of massive $\Phi$ particles to gauge fields  are only
string suppressed
\begin{equation}
\label{lambdaPhi}
  \lambda_{\Phi\gamma\gamma} \sim \frac{1}{m_s} \sim \frac{\sqrt{\mc{V}}}{M_p}
\end{equation}

The violent production of $\delta\tau_s$ fluctuations at the end of inflation leads to a huge number of canonical $\Phi$ particles
near the minimum, in addition to a small admixture of $\chi$ which arise due to the nontrivial mixing of $\delta \tau_s$ and $\delta\tau_b$, see equation (\ref{massmatrixmixing}).
The massive $\Phi$ particles subsequently decay to photons via perturbative processes such as $\Phi \rightarrow \gamma\gamma$ mediated
by the couplings in (\ref{canbasis}).  The decay width was computed in \cite{CQastro}
\begin{equation}
  \Gamma_{\Phi\rightarrow\gamma\gamma} = \frac{\lambda^2_{\Phi\gamma\gamma} m_\Phi^3}{64\pi}
\end{equation}
Using the estimates (\ref{mPhi}) and (\ref{lambdaPhi}) we find
\begin{equation}
\label{gamma_gg}
  \Gamma_{\Phi\rightarrow\gamma\gamma} \sim \frac{(\ln\mc{V})^3}{\mc{V}^2} M_p
\end{equation}

Also notice that, due to the $\delta\tau_b-\delta\tau_s$ mixing, a certain amount of $\chi$ particles will be produced from the oscillating inflaton field.  This does not lead to a cosmological moduli
problem because the $\chi$ particles are extremely massive for the value of $\mc{V}$ which we consider.  For example, taking $\mc{V} \sim 10^6-10^9$ we have $m_{\chi} \sim 10^4-10^8 \, \mathrm{GeV}$ which is
sufficiently massive that the residual $\chi$ moduli will decay well before Big Bang Nucleosynthesis (BBN) \cite{BBN}.\footnote{For very large values
of the compactification volume, $\mc{V} \sim 10^9$, the $\chi$ modulus may decay very close to the onset on BBN.  For such parameters a more
careful treatment of the decay of $\chi$, taking into account factors of order unity, may be necessary.}

\subsection{Inflaton Coupling to Fermions}

The inflaton $\tau_s$ will couple not only to photons but also to fermion fields living
on the world-volume of the D7 brane wrapping the inflationary 4-cycle.  The effective action
describing these interactions generically takes the form \cite{temp}:
\begin{equation}
  \mathcal{L}_{\mathrm{int}} = h_{\Phi\Psi\Psi} \Phi \, \bar{\Psi} \Psi
\end{equation}
where $h_{\Phi\Psi\Psi}$ is the dimensionless coupling and $\Psi$ schematically denotes any MSSM 
fermion.  The decay rate for this type of interaction is 
\begin{equation}
  \Gamma_{\Phi\rightarrow\Psi\Psi} \cong \frac{h^2}{8\pi} m_\Phi
\end{equation}
(in the limit $m_\Phi \gg m_\Psi$).  

Prior to electro-weak symmetry breaking a direct coupling to SM fields (such as the electron) is absent and the decay of 
$\Phi$ into fermions is dominated by the production of Higgsinos $\Phi \rightarrow\tilde{H}\tilde{H}$ and gauginos 
$\Phi \rightarrow \lambda\lambda$.  The relevant couplings were computed in \cite{temp}
\begin{eqnarray}
  h_{\Phi\tilde{H}\tilde{H}} &\sim& \frac{1}{\mc{V}^{1/2} \ln \mc{V}}  \\
  h_{\Phi\lambda\lambda} &\sim& \frac{1}{\mc{V}^{3/2} \ln \mc{V}} 
\end{eqnarray}
We therefore have the following estimates for the decay rates
\begin{eqnarray}
  \Gamma_{\Phi\rightarrow\tilde{H}\tilde{H}} &\sim& \frac{1}{\mc{V}^2 \ln \mc{V}} M_p \label{higgsino}\\
  \Gamma_{\Phi\rightarrow\lambda\lambda} &\sim& \frac{1}{\mc{V}^4 \ln \mc{V}} M_p
\end{eqnarray}
We see that the decays of $\Phi$ into Higgsinos dominate over the decays into gauginos.

\subsection{Reheating Temperature}

Let us now estimate the reheating temperature for the scenario where the SM lives on a D7 wrapping the inflationary
cycle $T_2$.  Note that $\Gamma_{\Phi \rightarrow \gamma\gamma} / \Gamma_{\Phi \rightarrow \tilde{H}\tilde{H}} \sim (\ln \mc{V})^4 \sim 10^4$ for a
compactification volume of order $\mc{V} \sim 10^6-10^9$, which is of interest for Roulette inflation.  Thus, comparing equations (\ref{gamma_gg})
and (\ref{higgsino}) we see that the lifetime of the inflaton is dominated by its decay into photons.  Using the standard result from the theory of 
reheating \cite{KLS97} the reheat temperature is given by
\begin{equation}
\label{Tr_wrapped}
  T_r \sim 0.1 \sqrt{\Gamma M_p} \sim 0.1 \frac{\left( \ln \mc{V}\right)^{3/2}}{\mc{V}} M_p
\end{equation}
Using $\mc{V} \sim 10^{6}$ we obtain $T_r \sim 10^{13}\, \mathrm{GeV}$ and using $\mc{V} \sim 10^9$ we have $T_r \sim 10^{10}\,\mathrm{GeV}$.  
Normally such a high reheat temperature would lead to a gravitino problem (specifically the late decays of the gravitino interfere with the successful 
predictions of Big Bang Nucleosynthesis).  However, in our case the gravitino mass is extremely high \cite{CQastro}:
\begin{equation}
  m_{3/2} \sim \frac{M_p}{\mc{V}}
\end{equation}
For $\mc{V} \sim 10^6-10^9$ we have $m_{3/2} \sim 10^{9}-10^{12}\,\mathrm{GeV}$, hence the gravitino is so massive that it decays well before nucleosynthesis \cite{BBN}.  
The disadvantage of such a high gravitino mass temperature is that SUSY is broken at too high a scale to explain the electro-weak hierarchy of the SM.  An alternative scenario to explain the fine-tuning 
of the Higgs mass is through vacuum selection effects on the string landscape \cite{splitSUSY}.  On the other hand, see \cite{newref} for a scenario where this gravitino mass is compatible with $\mathrm{TeV}$
scale soft terms.

There is, however, another cause for concern.  At sufficiently high temperatures thermal corrections to the effective potential tend to destabilize the moduli
and drive them to infinity, leading to decompactification \cite{temp}.  The decompactification temperature, $T_{\mathrm{max}}$, can be accurately approximated 
by $T_{\mathrm{max}} \cong V_b^{1/4}$ where $V_{b} \cong m_{3/2}^3 M_p$ is the height of the potential barrier separating the large volume AdS minimum from the supersymmetric minimum
at infinity in moduli space.  The maximum allowed reheat temperature (above which the internal space decompactifies) is therefore
\begin{equation}
\label{Tmax}
  T_{\mathrm{max}} \cong \frac{M_p}{\mc{V}^{3/4}}
\end{equation}

Comparing (\ref{Tr_wrapped}) to (\ref{Tmax}) one sees that, keeping track only of factors of the total volume $\mc{V}$, we have $T_r$ is marginally below the decompactification temperature.
However, this conclusion could easily be altered by factors of order unity which we have not taken into account.  Such factors depend sensitively on the details of the compactification
and we leave a careful determination to future studies.  Indeed, the requirement $T_r < T_{\mathrm{max}}$ may significantly constraint the parameter space of the model \cite{temp}.  

\section{Transfer of Energy into the SM Sector: D7 Wrapping a Non-Inflationary 4-Cycle}\label{sec:SM2}

As discussed previously, the wrapped D7 scenario of the previous section may lead to dangerous $g_s$-corrections to the inflaton potential.  Thus, it 
may be desirable to exclude such a wrapping.  Doing so, of course, also forbids a direct coupling between the inflaton and the SM fields.  In this case 
the inflaton may still decay to SM particles via some intermediate bulk states. Such decays which involve bulk modes must necessarily be 
suppressed by the compactification volume, which is exponentially large.  Here we discuss one possible scenario of this type.  

We could imagine the SM living on a D7 which wraps the big cycle $T_1$, however, this leads to unnaturally small gauge couplings \cite{CQastro}
and hence is phenomenologically disfavoured.  Instead, let us suppose the SM lives on a D7 wrapping some stabilized non-inflationary cycle, 
$T_3$, which is stabilized to a value $\langle\tau_3\rangle = \mathcal{O}(1)$.  This wrapping is illustrated schematically in Fig.~\ref{fig:CY} as ``scenario 2''.

At this point we must extend the discussion from section \ref{subsec:pertphotons} to the case with more than two moduli fields, namely, instead of 2 fields $\tau_b$ ($=\tau_1$)
and $\tau_s$ ($=\tau_2$) (corresponding to canonical moduli $\chi$ and $\Phi$) we consider 3 fields $\tau_1$, $\tau_2$ and $\tau_3$ (corresponding to canonical moduli $\chi$, $\Phi_2$
and $\Phi_3$).

As we have seen, the endpoint of inflation is marked by the violent production of fluctuations $\delta \tau_2$ of the inflationary 4-cycle, $T_2$.  Near the minimum of the potential
these $\tau_2$ fluctuations mix with the total volume $\tau_1$ and the non-inflationary hole $\tau_3$ through an off-diagonal mass matrix similar to 
(\ref{massmatrixmixing}), except now there are three relevant moduli rather than two.  Diagonalizing the mass matrix one finds three canonical fields
$\chi$ and $\Phi_i$ ($i = 2,3$) which are schematically related to the 4-cycle volumes as
\begin{eqnarray}
  \delta\tau_1 &\sim& \mc{O}(\mc{V}^{2/3}) \, \chi + \sum_i \mc{O}(\mc{V}^{1/6})\, \Phi_i \\
  \delta\tau_i &\sim& \mc{O}(\mc{V}^{1/2}) \, \Phi_i + \mc{O}(1) \, \chi + \sum_{j\not= i} \mc{O}(\mc{V}^{-1/2}) \Phi_j 
  \hspace{3mm}\mathrm{where}\hspace{3mm} i=2,3 \label{Phi_i_mix}
\end{eqnarray}
so that $\chi$ is mostly the total volume $\delta\tau_1$ while $\Phi_2$, $\Phi_3$ are mostly the blow-up modes $\delta\tau_2$ and $\delta\tau_3$ respectively.
The modulus $\chi$ is light, having mass
\begin{equation}
  m_\chi \sim \frac{1}{\mc{V}^{3/2} (\ln \mc{V})^{1/2} } M_p
\end{equation}
while the moduli $\Phi_2$, $\Phi_3$ are heavier 
\begin{equation}
  m_{\Phi_2} \sim m_{\Phi_3} \sim \frac{\ln \mc{V}}{ \mc{V} } M_p
\end{equation}

These moduli couple to gauge fields living on the D7 wrapping $T_3$ via the interaction $\tau_3 F_{\mu\nu}F^{\mu\nu}$, which gives us
\begin{equation}
\mathcal{L}_{\mathrm{int}} = \frac{\lambda_{\chi\gamma\gamma}}{4}\, \chi \, F^{\mu\nu}F_{\mu\nu} 
  + \sum_{i=2,3} \frac{\lambda_{\Phi_i\gamma\gamma}}{4}\, \Phi_i \, F^{\mu\nu}F_{\mu\nu}
\end{equation}
using equations (\ref{pert_L}) and (\ref{Phi_i_mix}).  Physically, these couplings originate from the mixing between $\tau_2$ and $\tau_1$, $\tau_3$.
From equation (\ref{Phi_i_mix}) we can estimate the magnitude of the moduli couplings.
The largest coupling is, obviously, the one involving $\Phi_3$ since the SM D7 wraps the cycle $T_3$.  This coupling is set by the string scale
\begin{equation}
  \lambda_{\Phi_3\gamma\gamma} \sim \frac{1}{m_s} \sim \frac{\mc{V}^{1/2}}{M_p}
\end{equation}
On the other hand, the coupling to the large cycle is
\begin{equation}
\lambda_{\chi\gamma\gamma} \sim \frac{1}{M_p\, \ln \mc{V}}
\end{equation}
where the factor $M_p^{-1}$ comes from that fact that $\chi$ is a bulk mode and the factor of $(\ln \mc{V})^{-1}$ is nontrivial \cite{CQastro}.
The coupling between the inflationary 4-cycle $T_2$ and brane-bound SM gauge fields is 
\begin{equation}
\label{Phi2gammagamma}
  \lambda_{\Phi_2\gamma\gamma} \sim \frac{1}{M_p} \frac{1}{\mc{V}^{1/2}}
\end{equation}
This is even more than Planck suppressed.


Thus, the picture of reheating in this scenario is the following.  Preheating produces copious amounts of $\delta\tau_2$ fluctuations which lead to a 
large
 number of canonical $\Phi_2$ moduli near the minimum, plus a small admixture of light $\chi$ and heavy $\Phi_3$ moduli.  The $\Phi_3$ particles
are the first to decay, they produce brane-bound SM photons via the process $\Phi_3 \rightarrow \gamma\gamma$ with a rate identical to 
(\ref{gamma_gg}):
\begin{equation}
  \Gamma_{\Phi_3\rightarrow\gamma\gamma} \sim \frac{(\ln \mc{V})^3}{\mc{V}^2} M_p
\end{equation}
This decay, however, does not correspond to true reheating since at this point the energy density of the universe 
is dominated by the almost nonrelativistic\footnote{See subsection \ref{sec:lattice} for a discussion of the fact that the bulk of the
inflation fluctuations produced by preheating are nonrelativistic near the minimum.} $\Phi_2$ particles with matter dominated equation 
of state, rather than SM radiation.  
Thus, the SM radiation which is produced at by the decay of $\Phi_3$ is rapidly diluted, $\rho_{\mathrm{radn}} \sim a^{-4}$, relative 
to the nonrelativistic $\Phi_2$ particles whose energy density dilutes as: $\rho_{\Phi_2} \sim a^{-3}$.

\emph{True} reheating occurs when the $\Phi_2$ particles subsequently decay into SM states via the suppressed coupling 
(\ref{Phi2gammagamma}).  This decay
proceeds with rate:
\begin{equation}
\label{GammaPhi2gammagamma}
  \Gamma_{\Phi_2\rightarrow\gamma\gamma} \sim  \frac{m_{\Phi_2}^3}{M_p^2 \mc{V}} \sim \frac{(\ln \mc{V})^3}{\mc{V}^4} M_p
\end{equation}
It is the decay rate (\ref{GammaPhi2gammagamma}) which determines the reheat temperature of the universe:
\begin{equation}
\label{Tr_nowrapping}
  T_r \sim 0.1 \sqrt{\Gamma M_p} \sim 0.1 \frac{(\ln \mc{V})^{3/2}}{\mc{V}^{2}} M_p
\end{equation}
This is smaller than the result of the previous section, equation (\ref{Tr_wrapped}), by a factor of $\mc{V}^{-1}$ corresponding to the fact that
the $\Phi_2$ is a bulk mode in this set-up with suppressed coupling to brane-bound states (\ref{Phi2gammagamma}).  This
suppression is actually favorable since it keeps the reheat temperature (\ref{Tr_nowrapping}) well below the decompactification scale 
$T_{\mathrm{max}}$,
given by (\ref{Tmax}).  Taking $\mc{V}\sim 10^6$ we have $T_r \sim 10^{7}\, \mathrm{GeV}$ and taking $\mc{V} \sim 10^9$ we have 
$T_r \sim 10\, \mathrm{GeV}$.  Hence, for extremely large values of the compactification volume, $\mc{V} \sim 10^9$, the reheating temperature
is so low that it may be difficult to realize baryogenesis at the electroweak phase transition.  We leave this issue to future investigations.

In this scenario the gravitino decays too early in the history of the universe to interfere with BBN: for $\mc{V} \sim 10^6-10^9$ we have 
$m_{3/2} \sim 10^{9}-10^{12}\, \mathrm{GeV}$.  As previously, the residual volume modulus $\chi$ particles decay after reheating but before BBN.


The results of this section may also have some relevance for the scenario where the SM cycle is much smaller than the string scale, see \cite{newref}.  We leave a detailed investigation to future studies.

\section{Stringy Reheating via K\"ahler Moduli Shrinking}\label{sec:shrink}

In this section, we propose yet another reheating mechanism in K\"ahler moduli inflation models.  This mechanism can operate even in the absence of a D7 wrapping the inflationary 4-cycle.
Hence, in the ensuing text we suppose (as in section \ref{sec:SM2}) that the SM lives on a D7 wrapping $T_3$.

The inflationary dynamics of $\tau(t)$ corresponds to the shrinking of the 4-cycle associated with $T_2$.
If this inflationary cycle of the internal CY shrinks to a minimal size $\tau_m$ comparable with the string scale, supergravity description breaks down
and new, stringy degrees of freedom must kick in. A natural candidate for such degrees of freedom is winding modes: closed strings with a nonzero winding number with respect to the 
inflationary 4-cycle. Because of winding number conservation winding string are created in pairs with winding numbers of equal absolute value but having opposite signs. Two such closed 
strings can merge into a one with zero winding number, {\it i.e.} a string that can move away from the inflationary 4-cycle and into the bulk region of the CY. In other words, the energy
of the shrinking inflationary cycle ({\it i.e.} inflaton kinetic energy) is transferred into closed string excitations living in the bulk.  This is somewhat similar to the energy transfer from annihilating
branes into closed string excitations that occurs at the end of stringy warped throat brane inflation \cite{BBC,KY}.  The excited closed strings decay via many cascades into closed strings with 
the lowest level of excitation, {\it i.e.}, KK gravitons. The produced KK gravitons interact with the SM brane scalar moduli $Y_i$ (corresponding to the transverse fluctuations of the brane) and also 
SM particles.  Consider, for example, the interaction between KK gravitons and brane moduli $Y_i$.  This interaction is gravitational 
and can be described by a vertex $\frac{1}{M_P^2} h^{KK}_{\mu\nu} T^{\mu\nu}(Y)$, the corresponding decay width is 
\be\label{GKKgr}
\Gamma_{KK \to Y} \sim \frac{m^3_{KK}}{M_P^2}.
\ee  
As discussed in \cite{KY}, this coupling is sufficiently generic to work for any low-lying degrees of freedom on the SM brane world.  (In particular, world-volume fermions.)  
For the purposes of making a rough estimate of the reheat temperature we can safely assume that $\Gamma_{KK\to Y} \sim \Gamma_{KK \to SM}$.

Let us now estimate (\ref{GKKgr}).  The KK modes have typical mass $m_{KK} \sim 1/R_{CY}$, where the overall size of the internal CY is $R_{CY} \sim \mc{V}^{1/6}\; \sqrt{\alpha'}$. In the
models under consideration we have
\be
\sqrt{\alpha'} \sim \frac{1}{m_s} \sim \frac{\sqrt{\mc{V}}}{M_P},
\ee
so that $R_{CY} \sim \frac{\mc{V}^{2/3}}{M_P}$, $m_{KK} \sim \frac{M_P}{\mc{V}^{2/3}}$, and
from (\ref{GKKgr}) we have
\be\label{GKK}
\Gamma_{KK \to SM} \sim \frac{M_P}{\mc{V}^{2}}.
\ee
The decay rate (\ref{GKK}) is suppressed as compared to the case with the wrapped D7, equation (\ref{gamma_gg}) by a logarithm which gives a numerical factor of about $10^{-3}$.  

The decay rate (\ref{GKK}) allows us to estimate the reheat temperature as
\begin{equation}
  T_r \sim 0.1 \frac{1}{\mc{V}} M_p 
\end{equation}
Taking, for illustration, $\mc{V} \sim 10^6$ we have $T_r \sim 10^{11} \, \mathrm{GeV}$ and taking $\mc{V} \sim 10^9$ we have $T_r \sim 10^{8}\, \mathrm{GeV}$.  
As in the scenarios discussed in sections \ref{sec:SM1} and \ref{sec:SM2}, such a high reheat temperature is phenomenologically sensible because the gravitino is extremely massive.
As in the scenario discussed in section \ref{sec:SM2} (but unlike the scenario is section \ref{sec:SM1}) this reheat temperature is well below the 
decompactification scale, $T_r / T_{\mathrm{max}} < \mc{V}^{-1/4} \ll 1$.  

\section{Summary and Discussion}\label{sec:conc}

Modular inflation, and other string theory inflation models, provide a natural 
playground for studying reheating in a context where (at least in theory) one can actually determine all decay channels of the inflaton to the visible 
(SM) sector from first principles.  Here we investigated in detail reheating after modular/Roulette string theory inflation models in the context of large volume compactifications.
Our results show that in realistic microscopic models the details of reheating can be rather complicated and may proceed through a variety of channels, including perturbative 
decays, nonperturbative preheating and also inherently stringy processes.  

We found that modular inflation models (such as K\"ahler modulus or Roulette inflation) in particular display very rich post-inflationary dynamics.  
\begin{itemize}
  \item {\it The initial stages} of the decay of the inflaton proceed via extremely efficient nonperturbative  particle production involving a combination of tachyonic 
instability and parametric resonance.  This combination may lead to the most violent example of preheating known in the literature.  
\end{itemize}
The subsequent 
stages of reheating involve the transfer of energy from the inflaton into excitations of the visible  sector.  This phase is more model dependent since we 
must identify the location of the standard model of particle physics in the compactification volume.  We have considered three separate scenarios 
for transferring the energy from the inflaton fluctuations $\delta\tau$ into SM degrees of freedom.  These are as follows:
\begin{itemize}
  \item \emph{D7 wrapping $T_2$} (assuming a SUGRA description of reheating is valid, i.e., the minimum hole size is larger than the string length):
  In this case the SM fields live on a D7 wrapping the inflationary 4-cycle, $T_2$, so that there is a direct coupling 
  between the SM fields and the inflaton $\tau$.  Photon production (both perturbative and non-perturbative) dominates the decay of the inflaton.  
  The resulting reheat temperature is given by (\ref{Tr_wrapped}).  Using $\mc{V} \sim 10^{6}$ we obtain $T_r \sim 10^{13}\, \mathrm{GeV}$ and using 
  $\mc{V} \sim 10^9$ we have $T_r \sim 10^{10}\,\mathrm{GeV}$.  Although such reheat temperatures are extremely high, there is no problem with BBN because  
  $m_{3/2} \sim 10^{9}-10^{12}\,\mathrm{GeV}$ is so large that the gravitinos decay before the onset of nucleosynthesis.  However, such a high reheat temperature
  may lead to decompactification.  This scenario also has the disadvantage that $g_s$-corrections may spoil the flatness of the inflaton potential.  
  \item \emph{D7 wrapping $T_3$} (assuming a SUGRA description of reheating is valid, i.e., the minimum hole size is larger than the string length): 
  In this case the SM fields live on at D7 wrapping some non-inflationary 4-cycle, $T_3$ (we exclude
  the big cycle $T_1$ since this would give unacceptably small gauge couplings).  Near the minimum of the potential the produced inflaton 
  fluctuations $\delta\tau_2$ mix with the fluctuations of the big 4-cycle $\delta\tau_1$ and also of the small non-inflationary 4-cycle $\delta\tau_3$
  through a non-diagonal mass matrix.  The decay of the $\delta\tau_2$ fluctuations into brane-bound SM states via Planck suppressed operators 
  leads to reheating.  
  This scenario 
  evades any dangerous $g_s$-corrections to the inflaton potential and the fact that reheating proceeds through Planck suppressed operators 
  suppresses the reheat temperature to well below the decompactification scale.   Taking $\mc{V}\sim 10^6$ we have $T_r \sim 10^{7}\, \mathrm{GeV}$ and taking $\mc{V} \sim 10^9$ we have $T_r \sim 10\, \mathrm{GeV}$.
  As above, there is no gravitino problem because $m_{3/2} \sim 10^{9}-10^{12}\,\mathrm{GeV}$. 
  \item \emph{Stringy reheating}: In this case closed strings are produced when the inflationary 4-cycle size becomes of order the string length.  These closed strings 
  cascade into KK gravitons which can subsequently interact with brane-bound SM fields (wrapping some non-inflationary 4-cycle) via Planck-suppressed operators.  Taking, for illustration, $\mc{V} \sim 10^6$ we have 
  $T_r \sim 10^{11} \, \mathrm{GeV}$ and taking $\mc{V} \sim 10^9$ we have $T_r \sim 10^{8}\, \mathrm{GeV}$. There is, again, no gravitino problem 
  because $m_{3/2}$ is so large. Decompactification is evaded due to the Planck suppression
  of the interactions between bulk and brane modes.  This scenario also evades potentially dangerous $g_s$-corrections to the inflaton potential.
\end{itemize}
For convenience we summarize the rates for the dominant decay channel of the inflaton in the various scenarios in the following table:
\vspace{2mm}
\begin{center}
\begin{tabular}{|c|c|}
\hline
\multicolumn{2}{|c|}{{\bf Summary of Dominant Decay Rates}} \\ \hline
$\Gamma / M_p$ & Scenario \\ \hline
  & \\
$\sim \frac{(\ln \mc{V})^3}{\mc{V}^2} $  & D7 wrapping $T_2$\\
  & \\
$\sim \frac{(\ln \mc{V})^3}{\mc{V}^4} $  & D7 wrapping $T_3$\\
  & \\
$\sim \frac{1}{\mc{V}^2}$  &  stringy reheating from shrinking hole \\
  & \\
\hline
\end{tabular}
\end{center}
\hspace{2mm}
In all cases considered there is the disadvantage that SUSY is broken at too high a scale to explain the small value of the Higgs mass.  This happens 
because the typical compactification volumes which are favorable for inflation $\mc{V} \sim 10^6 - 10^9$ are much smaller than the value 
$\mc{V} \sim 10^{15}$ that is favored for particle phenomenology \cite{CQastro}.\footnote{This tension is similar to what happens in KKLMMT brane inflation 
where the value of the warping that would be required to solve the hierarchy problem \emph{\`a la} Randall and Sundrum is much larger than the warping 
that is favorable for inflation.  In the case of KKLMMT this can be evaded by adding additional throats to the compactification \cite{BBC} at the expense 
of complicating the reheating process.}  See, however, \cite{newref} for a scenario where $\mc{V} \sim 10^6-10^7$ may be compatible with $\mathrm{TeV}$
soft terms.

Our first scenario, the SM D7 wrapping $T_2$, is afflicted by two potential complications: $g_s$-corrections may destroy the flatness of the inflaton potential and the reheat temperature is so high that it may lead to decompactification.  
Both of these issues will require further study, however, at first glance this scenario seems disfavoured.  On the other hand, the scenario with the SM D7 wrapping $T_3$ evades both of these potential difficulties and therefore
seem most promising.   For such models  there are again a number of directions open for further study.  
Also, it would be interesting to perform a more detailed study the topology-changing transition in the case where the inflationary hole size shrinks to the string length at the end of inflation.  
We leave these, and other, interesting issues to future investigations.

\section*{Acknowledgments}

We are grateful to M.\ Cicoli, J.\ Cline, J.\ Conlon, K.\ Dasgupta, R.\ Kallosh, A.\ Linde and F.\ Quevedo for helpful discussions and comments.  
Thanks also to P.\ Vaudrevange and S.\ Prokushkin for early collaboration.  This work was supported in part by NSERC.

\end{document}